\documentstyle[12pt,epsfig]{article} 
 \newcommand {\bi} {\bibitem}
 \newcommand {\be} {\begin{equation}}
\newcommand {\bea} {\begin{eqnarray} \nonumber }
\newcommand {\ee} {\end{equation}}
\newcommand {\eea} {\end{eqnarray}}
 \newcommand {\eps} {\epsilon}

\newcommand {\la} {\lambda}

 \newcommand {\al} {\alpha}

\newcommand {\Tr} {\mbox{Tr}}

\def \form#1 {eq. (\ref{#1}) }
\def \parziale#1#2  {{\partial {#1} \over \partial {#2}}}

\def\bW1{{\bf  W_1}}
\def\bh1{{\bf  h_1}}

\def\(({\left(}
\def\)){\right)}
\def\[[{\left[}
\def\]]{\right]}

\def \la{\langle}
\def \ra{\rangle}
\def \intk{\int {d^dk \over (2 \pi)^3}}
\def \eps{\epsilon}

\newcommand{\nn}{\nonumber}

\begin{document}   
\title{A first principle computation of the thermodynamics
%C
 of glasses}
\author{ 
Marc M\'ezard \\
{\it Institute for Theoretical Physics, University of California Santa Barbara}
\\ 
{\it CA 93106-4030, USA, and LPTENS, CNRS, France}
\\and\\
 Giorgio Parisi 
 \\
{\it Dipartimento di Fisica,
Universit\`a {\sl La Sapienza} and INFN Sezione di Roma I} \\
\it{ Piazzale Aldo Moro, 
Roma 00187, Italy}
}
\maketitle
\begin{abstract}
We propose a first principle computation of the equilibrium thermodynamics of 
simple fragile glasses 
starting from the two body interatomic potential.  A replica formulation 
translates this problem 
into that of a gas of interacting molecules, each molecule being built of $m$ 
atoms, and having a 
gyration radius (related to the cage size) which vanishes at zero temperature.
 
We use a small cage 
expansion, valid at low temperatures, which allows to compute the cage size,
the 
specific heat 
(which follows the Dulong and Petit law), and the configurational entropy.  
\end{abstract}
\newpage
\section{Introduction}
Take a three dimensional classical system consisting of many particles, 
interacting through a short 
range potential with a repulsive core.  Very often this system will undergo, 
upon cooling or upon
%C 
compression, a solidification into an amorphous solid state - the glass 
state.  The 
conditions required for observing this glass phase is the avoidance of 
crystallisation, which can 
always be obtained through a fast enough quench (the meaning of 'fast' depends 
very much of the type 
of system) \cite{glass_revue}.  There also exist potentials which naturally 
present some kind of 
frustration with respect to the crystalline structure.
%C
 Whether their actual
stable state is a crystal or a glass is not known, but they are
known to  solidify 
into glass states, 
even when cooled slowly - such is the case for instance of binary mixtures of 
hard spheres, soft 
spheres, or Lennard-Jones particles with appropriately different radii.  These 
have been studied a 
lot in recent numerical simulations
\cite{gpglass,BK1,BK2,FRAPA,ColPar}.

Our aim is to compute the equilibrium thermodynamic properties of this glass
phase, using 
the statistical 
mechanical approach, namely starting from the microscopic Hamiltonian
(an attempt to build up a non equilibrium thermodynamic phenomenology can be 
found
in \cite{theo}).  We shall 
therefore assume 
that crystallisation has been avoided, and consider only the amorphous solid 
state.  The general 
framework of our approach finds its roots in old ideas of Kauzman 
\cite{kauzman}, Adam and Gibbs
\cite{AdGibbs}, which received a boost when Kirkpatrick, Thirumalai and
Wolynes 
underlined the 
analogy between structural glasses and some generalized spin glasses 
\cite{KiThiWo}.  
%C
 This 
framework should provide a good description of fragile glass-formers.
These are the systems in which the increase of relaxation time upon 
decreasing the temperature is much faster than Arrhenius- often
parametrized as a Vogel-Fulcher law, displaying a divergence
of the relaxation time at a finite temperature\cite{glass_revue}.
 In this approach the 
glass transition, 
measured from dynamical effects, is supposed to be 
associated with an underlying thermodynamic 
transition at the 
Kauzman or Vogel-Fulcher temperature $T_K$.  This ideal glass transition is
the 
one which should be 
observed on infinitely long time scales in
fragile glass-formers\cite{glass_revue}.  This transition is 
of an unusual type, 
since it presents two apparently contradictory features: 
\begin{enumerate} 
\item The transition is continuous (second order) from the thermodynamical
point 
of view: the internal energy 
is continuous, and the transition is signalled by a discontinuity of the 
specific heat which jumps 
from its liquid value above $T_K$ to a value very close to that of a crystal 
phase below.
\item The order parameter is discontinuous at the transition.
\end{enumerate}
 
In order to make this last statement precise we shall have to define an order 
parameter for the 
glass phase in the framework of equilibrium statistical mechanics, which 
involves some subtleties 
and will be addressed below.  At this introductory stage let us take loosely
as 
an order parameter 
the correlation in the positions of the particles at very large times.  In the 
liquid there is no 
correlation.  In the glass the positions are correlated in time.  Clearly the 
order parameter jumps 
discontinuously between the liquid phase and the glass phase.  The two 
properties above are indeed 
observed in generalized spin glasses
\cite{GrossMez}. The problem of the existence or not of a diverging correlation
length is still an open one \cite{corr_length}.

This analogy is suggestive, but it also hides some very basic differences,
like 
the fact that spin 
glasses have quenched disorder while structural glasses do not.  The recent 
discovery of some 
generalized spin glass systems without quenched disorder 
\cite{nodis1,nodis2,nodis3} has given 
credit to the idea that this analogy is not fortuitous.  The problem was to
turn 
this general idea 
into a consistent computational scheme allowing for some quantitative 
predictions.  Important steps 
in this direction were invented in \cite{remi,pot}, which showed how useful it 
is to study several 
coupled copies of the same system in order to characterize properly the glass 
phase.  In a previous 
preliminary study, we used some of these ideas to estimate the glass 
temperature, arriving from the 
liquid phase
\cite{MPhnc}.  
However the approximations we did were not adequate for the description of the 
low temperature 
phase.  Here we concentrate instead on the properties of the glass phase
itself 
and we introduce 
approximations which are much more appropriate to describe its properties 
particularly at low 
temperatures.  We are now able to construct analytical tools for doing 
computations in the glass 
phase and to test the results in numerical (and eventually real) experiments.
A brief description of a part of the present work has appeared in 
\cite{MPglass1}.

In the next section we shall present in more details the general physical 
picture underlying
our approach.
In sect.3 we shall explain why and how one can characterize and study the
glass 
phase using a
replicated liquid. Sect. 4 derives the Hamiltonian of the molecular liquid,
which is studied in the next two sections, first of all by a small cage 
expansion in sect.5, then by a molecular HNC closure in sect.6.  In sect.7
we present the results of these various approximations concerning the glass
transition temperature and the  thermodynamic quantities. Sect. 8 
gives a list of some  directions into which this work could be extended. Two 
appendices
contain the derivation of the molecular HNC closure on one hand, and its 
expansion to
second order in the small cage parameter on the other hand.

\section{The basic scenario}

In this section we want to present some of the general ideas which
provide a background to our approach. These have to do with the
existence of a configurational entropy, and the identification
of the glass transition as a point where the configurational entropy 
vanishes. These ideas are presented in general, without special
reference to a specific system. They can be derived in great details in some
mean field spin glass models. Although the microscopic description of
these models is somewhat remote 
from the actual glass problem which interests us, we have included  for 
completeness a short summary of some of the results found in these
systems. This will help to formulate the basic hypotheses of
our approach.

\subsection{Configurational entropy}
We consider 
a system of $N$ particles moving in a volume $V$ of a d-dimensional space,
and interacting by some short range potential. These could be for instance hard
spheres or Lennard-Jones particles.

\begin{figure}
\centerline{\hbox{
\epsfig{figure=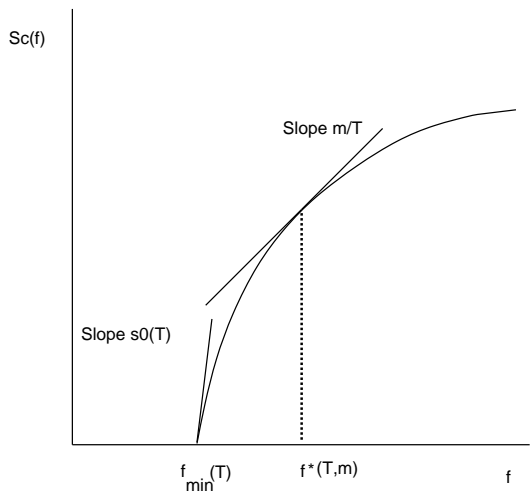,width=7 cm,angle=0}
}}
\caption{Qualitative shape of the configurational entropy versus free energy. 
The
whole curve depends on the temperature. The
saddle point which dominates the partition function,
for $m$ constrained replicas, is the point $f^*$ such
that the slope of the curve equals $m/T$ (for the usual unreplicated system, 
$m=1$).
If the temperature is small enough the saddle point sticks
to the minimum $f=f_{min}$ and the system is in its glass phase. }
\label{sigma_qualit}
\end{figure}

Let us introduce the  free energy functional 
$F[\rho]$ 
which depends on the local particle density $\rho(x)$ and on the temperature.  
We suppose that at sufficiently low 
temperature this functional has an exponentially large number of minima
\cite{still}. More precisely, the number of free energy minima with 
free energy density  $f=F/N$ is supposed to be exponentially large in 
some region of free energies, $f_{min}(T)<f<f_{max}(T)$:
\be
{\cal N}(f,T,N) \approx \exp(NS_c(f,T)) \ .
\label{CON}
\ee
Exactly at zero temperature these minima coincide with the mimima of the 
potential energy as 
function of the coordinates of the particles.  The function $S_c$ is called
the 
complexity or the 
configurational entropy (it is the contribution to the entropy coming from the 
existence of an 
exponentially large number of locally stable configurations).  The number of 
local minima is 
supposed to vanish outside of the region $f_{min}(T)<f<f_{max}(T)$, and the 
configurational entropy 
$S_c (f,T)$ is supposed to go to zero continuously at $f_{min}(T)$, as found
in 
all existing models 
so far (see fig.\ref{sigma_qualit}).

Let us first discuss the properties of the system at thermal equilibrium: we 
thus consider the case 
where each configuration of the system is assigned a probability given by its 
Boltzmann weight.  We 
label the free energy minima by an index $\alpha$.  To each of them we can 
associate a free energy 
$F_\al$ and a free energy density $f_\al= F_\al/N$.  In the low temperature 
region we suppose that 
the total free energy of the system ($\Phi$) can be well approximated by the
sum 
of the 
contributions to the free energy of each particular minimum:
\be
Z\equiv \exp(-\beta N \Phi) \simeq\sum_\al \exp(-\beta N f_\al).
\ee
For large values of $N$ we can write
\be
\exp(-N \beta \Phi) \approx \int_{f_{min}}^{f_{max}} df \exp [-N(\beta f- 
S_c(f,T))] \ .\label{SUM}
\ee
We can thus use the saddle point method and  approximate the 
integral  with the integrand evaluated at its maximum.
We find that
\be
\Phi=\min_f\Phi(f) \equiv  f^* -  T S_c(f^*,T),
\ee
where
\be
\Phi(f)\equiv f - T S_c(f,T).
\ee
This formula is quite similar to the usual formula for the free energy ,i.e.  
$f=\min_{E} ( E - T S(E))$, where $S(E)$ is the entropy density as 
a function of the energy density ($E$).
The main difference is the fact that the total entropy of the 
system has been decomposed into the
contribution due to small fluctuations around a given configuration (this
piece 
has
been included into $f$), and the contribution due to the existence of a large 
number of
locally stable configurations, the configurational entropy.

Calling  $f^*$ the 
value of $f$ which minimize $\Phi(f)$, we have two possibilities:
\begin{itemize}
\item
The minimum lies inside the interval and it can be found as the solution 
of the equation $\beta=\partial S_c/\partial f$.  In this case we have
\be
 \Phi= f^* - T S_c^*, \ \ \ S_c^*=S_c(f^*,T).
\ee
The system may stay in one of the many possible minima.  The number of 
accessible minima 
 is $\exp(N S_c^*)$ .  The entropy of the system is thus 
the 
sum of the entropy of a typical minimum and of $S_c^*$, which is the 
contribution to the entropy 
coming from the exponentially large number of metastable configurations.

\item
The minimum is at the extreme value of  the range of variability of 
$f$: it sticks at $f^*=f_{min}$ and the total free energy is
$\Phi=f_{min}$.  In this case the contribution of the configurational entropy
to 
the 
free energy is zero.  The different states which contribute to the free energy 
have a difference in 
free energy density which is of order $N^{-1}$ (a difference in total free 
energy of order 1).
This situation is often encountered in spin glasses, both in usual cases
of spin glasses with quenched disorder \cite{MPV,parisibook2}, and also in
some spin glass systems without quenched disorder \cite{nodis1,nodis2,nodis3}.

\end{itemize}

One aim of the theory of glasses at equilibrium could be to demonstrate from 
first principles the 
existence of a configurational entropy function such as depicted in fig.  
\ref{sigma_qualit}, 
and to compute it.  
This is difficult to achieve.  For instance Kepler's conjecture, a simple zero 
temperature statement 
saying that there is no denser packing of hard spheres in three dimensions
than 
the fcc lattice, has 
resisted a proof for more than three centuries \cite{kepler}.  Here we shall 
take a 
more modest starting 
point: we shall assume the existence of the local minima and of the 
configurational entropy 
function with the 
general properties depicted above, and within this assumption we shall show
how 
to compute 
(approximately but with a rather good accuracy, and one which can be improved 
systematically) the 
various properties of the system, including the configurational entropy
function 
itself.

\subsection{Mean field situation}
So far, the only systems for which the above program
 could be carried out
in all details are some type of mean field spin glasses with a discontinuous
jump of the order parameter at the transition 
\cite{GrossMez,KiThiWo,crisomtap,kurparvir,ACP,MMpspin}.

Although we will not need all the ingredients that have been found in these 
other problems, it is 
useful to recall some of them; later on we will mention how this picture might 
be modified in a 
realistic - non mean field - system.  The configurational entropy function is 
convex, and 
previous work indicates 
that it depends smoothly on the temperature, the main effect of a temperature 
change being a global 
shift of the free energies.  Starting from high temperatures, we thus
encounter 
the following 
temperature regions (we use here the language of liquids and glasses).

\begin{itemize}
\item For $T>T_D$ the free energy functional is dominated  by the uniform 
density
solution, $\rho(x)=\rho$ (there may exist close to $T_D$ other minima
 \cite{BaBuMez,kurparvir}, but their
total contribution has a higher free energy than the uniform solution).
The system is obviously in the fluid phase.

\item In the region where $T_D>T>T_K$, the minimum of the function $\Phi(f)$
is 
within
the interval $[ f_{min}(T), f_{max}(T) ]$. Therefore the system can stay in
one 
of many
different states. The entropy of the equilibrium system receives a contribution
from the configurational entropy, which is non zero. A very surprising result, 
found in all generalized mean field spin glasses with discontinuous 
transition so far, is that the 
total free energy of the system including the configurational entropy 
contribution, 
$\Phi(f^*)$,
is {\it equal} to the free energy of  the fluid solution with uniform $\rho$
\cite{remi,pot}. This result has not received a general explanation beyond 
the
simple idea of the transition at $T_D$ being a fragmentation of accessible phase
space into many separated pockets, the total volume of which  should
be continuous at $T_D$.
Although the  thermodynamics is still given by the usual expressions of the 
liquid phase
 and the 
free energy is analytic at $T_D$, below this temperature the  system, 
 at each given moment of time, may stay in one of the exponentially large
number 
of 
minima.
 
 \item
In the region where $T<T_K$ the saddle point of $\Phi$ sticks at its minimum
and the free energy is dominated by the contribution of a few minima 
 having the lowest possible value $f_{min}(T)$.  Here the free energy is no
more 
the analytic 
continuation of the free energy in the fluid phase.  A phase transition is 
present at $T_K$ and the 
specific heat is discontinuous here. 
\end{itemize}

\begin{figure}
\centerline{\epsfxsize=7cm
\epsffile{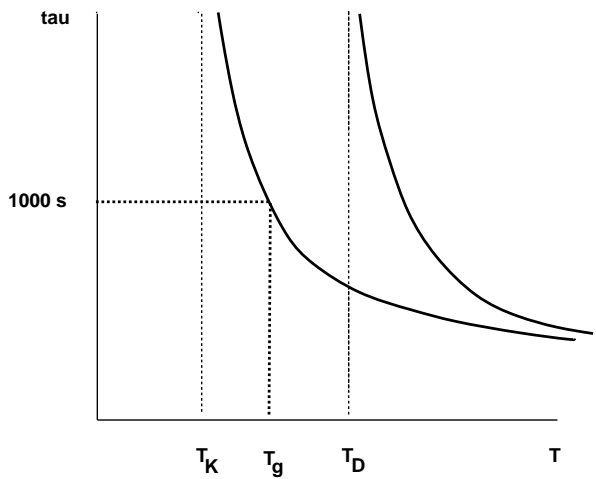}
}
\caption{Relaxation time {\it vs} temperature in discontinuous spin-glasses.
 The right hand curve is the mean-field
prediction, which gives a dynamical transition at a temperature $T_D$ above
the static transition temperature $T_K$. The left curve is a conjecture on the
behaviour
in finite dimensional systems: activated processes smear the dynamic
transition. The
relaxation time diverges only at the static temperature $T_K$, but becomes
experimentally
large already around the glass temperature $T_g$. }
\label{CritTemp}
\end{figure}

The intermediate phase $T_D>T>T_K$ is particularly interesting.  In the mean 
field systems, an exact 
solution of the Langevin dynamics indicates a dynamical phase transition at 
$T_D$, the system being 
trapped in some states with a free energy which is extensively higher than
that 
of the equilibrium 
state \cite{cuku}.  For the realistic finite dimensional problems which we
want 
to study,
%C
the situation is much less  clear, but one can speculate  
that the system will equilibrate in this regime,  very slowly 
\cite{KiThiWo}.  The time 
to jump from one minimum to another minimum is quite large: it is an activated 
process which is 
controlled by the height of the barriers which separate the different minima.  
The correlation time 
will become very large below $T_D$ and for this reason $T_D$ is called the 
dynamical transition 
point.  The correlation time (which should be proportional to the viscosity) 
diverges only at the 
true thermodynamic transition temperature, sometimes called the ideal glass 
temperature $T_{K}$ (see 
fig.\ref{CritTemp}).
  The precise form of  this divergence is not well 
understood.
It is natural to suppose that one should get a divergence of the form 
$\exp(A/(T-T_{K})^{\nu})$ for 
an 
appropriate value of $\nu$, but a reliable analytic computation of $\nu$ is 
lacking 
\cite{KiThiWo,PAK}.
Experiments can often be fitted by this law with various values of $\nu$, 
including the 
Vogel-Fulcher fit with $\nu=1$.  The equilibrium configurational entropy is 
different from zero (and 
it is a number of order 1) when the temperature is smaller than $T_D$, it 
decreases with the 
temperature and it vanishes linearly at $T=T_K$.  At this temperature the 
entropy of a single 
minimum becomes equal to the total entropy and the contribution of the 
configurational entropy to 
the total free energy vanishes.  Therefore the total entropy and the internal 
energy are continuous 
at the transition.

\subsection{Relationship to experiments}
The above scenario is appealing in that it puts into a unified framework a 
number
of experimental facts on glasses, as well as some general theoretical ideas.

Experimentally, the system falls out of equilibrium when
its relaxation time becomes larger than the experimental time. The `glass
transition temperature', defined conventionally as the temperature where the 
typical
relaxation time reaches a value of order one hour, falls somewhere between 
$T_K$ and $T_D$. By considering slower and slower quenches, one can
equilibrate 
the
system at lower temperatures. However in this scenario there exists an 
underlying
thermodynamic transition at the temperature $T_K$, which is the ideal glass 
transition
temperature. This temperature is also the one where the viscosity would
diverge   in the Vogel-Fulcher type fitting
of the viscosity versus temperature. Clearly it also corresponds to
the Kauzman temperature: the excess entropy of the supercooled liquid with 
respect
to the crystal is basically equal to the configurational entropy, which vanishes
precisely at $T_K$. The experimental fact that the Kauzman temperature and 
the Vogel-Fulcher one 
are close to each other has been noted many times, and is also found in the
Adam-Gibbs relation \cite{AdGibbs}. 

The dynamical temperature $T_D$ also
receives a natural interpretation. In mean field, therefore
neglecting activated processes, the relaxation time diverges with a power
law at $T_D$, and the autocorrelation function develops an infinitely
long plateau. This  slowing down is  described precisely by the mode coupling
theory 
\cite{BCKM}.
In the mean field approximation the height of the barriers separating the 
different minima is 
infinite and the temperature $T_D$ is sharply defined as the point where the 
correlation time 
diverges.  In the real world activated process (which are neglected in the
mean 
field approximation 
and consequently in the mode coupling theory) have the effect of producing a 
finite (but large) 
correlation time also at and below $T_{D}$ (the  precise meaning of 
the dynamical 
temperature beyond mean 
field approximation is not so clear -see \cite{FRAPA}; probably the best 
definition is
that $T_D$ is the temperature where the mode-coupling theory predicts a 
transition).
 Therefore one expects
that the mode coupling description will give good results in the region
largely 
above
$T_D$, a fact that has been checked accurately in  experiments \cite{MCT_exp}
and numerical simulations\cite{Kob}.

A last point which is predicted within the basic scenario, and has been
checked numerically, is a specific type of aging and modification
of the fluctuation-dissipation relation.
%C
The aging behaviour, which has been seen many years ago already 
in some polymeric
glasses \cite{POLI}, can be studied in details in spin glasses
\cite{BCKM_rev}. 
These studies, initiated by the work of Cugliandolo and Kurchan \cite{cuku},
have led to some well defined generalisation of the basic equilibrium properties
such as  time translation invariance and fluctuation-dissipation theorem
(FDT). 
This 
generalisation is not limited to the narrow scope of some special mean field
spin-glasses, but  seems to provide a general description of glassy dynamics
in 
many
systems, including structural glasses.
  The modification of the fluctuation-dissipation relation 
  can be measured directly, although the experiments are 
not simple.  On the 
other hand, numerical simulations for a binary mixture of soft spheres
\cite{gpglass} or Lennard-Jones particles \cite{BK2} have found exactly
the non-trivial modification which is predicted by the general scenario, 
providing
therefore a confirmation of its validity at least on their (limited)
time scales.

\section{A static order parameter for the glass phase}
\label{strategy}
In this section we wish to explain the general strategy for describing and
computing properties of an amorphous solid state. We are
particularly interested in
 systems with many metastable states, having a non zero configurational entropy.
We shall explain the general strategy trying to keep away as much as possible 
from 
any specific model, the more precise formulation for our problem will be
given in the next section.
 Let us 
consider 
a system of $N$ particles, interacting  by a two body potential with a 
Hamiltonian
\be
H=\sum_{1 \le i \leq j \le N} v(x_i-x_j)
\ee
where the particles move in a volume $V$ of a d-dimensional space, and $v$ is an
arbitrary short range potential with a short  range repulsion, like 
a $1/r^{12}$ potential or a Lennard-Jones one. We shall
 take the thermodynamic limit $N,V \to \infty$ at fixed density 
$\rho=N/V$. For simplicity,  we do not consider here the description of
mixtures 
of different
types of particles. The generalization to mixtures is necessary if one wants
to 
compare
more precisely to simulations, which are performed on mixtures in order to
avoid 
crystallisation. 
This generalization, together with a detailed comparison, will be presented in a
forthcoming paper \cite{BPGM}. Some general background is provided by
the review paper \cite{BCKM_rev}.

\subsection{Time persistent correlations}
Before going to a purely static description of the order parameter, let us
first 
discuss a dynamical one. At an atomic level one often tends to associate the 
glass transition
with the divergence of the time scale on which a labelled particle can get out 
of its trap. While this is an intuitive picture, it is not possible to
translate 
it 
into
a good definition of the solid phase: because of the  excitation and movements 
 of vacancies and other defects,
this individual trapping time scale is always finite, although it will
increase 
exponentially
when the temperature gets small. In order to get a proper definition of the 
solid,
it has been proposed \cite{Geszti,goldbart} to use a generalisation of the 
Edwards 
Anderson order
parameter of the type:
\be
Q_{EA}(p)=\lim_{t \to \infty} \lim_{N \to \infty} {1 \over N} \sum_{jk} \la
e^{ip \cdot (x_j(0)-x_k(t))}\ra
\ee
where $p$ is an arbitrary non zero wave vector, the order of magnitude of
which 
is one over the typical interparticle distance.  When the system is in the 
liquid phase 
the above order 
parameter is zero and when it is in the glass phase this order parameter is
non 
zero (even in the 
presence of single particle diffusion).

This definition would hold for the equilibrium dynamics, i.e. assuming that
the 
system is
in equilibrium at time $t=0$. As we know the glass never reaches equilibrium
and 
therefore
it ages: correlations are not stationary in time. The proper generalization of 
the
previous correlation taking into account the aging effect takes the slightly 
more complicated form
(where the order of limits is crucial):
\be
Q_{EA}(p)=\lim_{\tau \to \infty} \lim_{t_w \to \infty} \lim_{N \to \infty}
 {1 \over N} \sum_{jk} \la
e^{ip \cdot (x_j(t_w)-x_k(t_w+\tau))}\ra
\label{EAvrai}
\ee

This gives a sensible dynamical definition of the glass phase.

\subsection{Correlations between two copies}

We would like a purely static description of the solid phase in the framework
of 
equilibrium 
statistical mechanics, in a case where there are no Bragg peaks.  As soon as
we 
have a solid phase 
the translational symmetry is broken and the system can be in many states. 
For 
crystalline order 
these many states just differ from each other by rotations or translations
which 
can be easily taken 
care of by appropriate boundary terms.  In the glass case, in order to choose
a 
state, one should 
first know the average position of each atom in the solid, which requires an 
infinite amount of 
information.  Had we known this information, we could have added to the 
Hamiltonian an infinitesimal 
but extensive pinning field which attracts each particle to its equilibrium 
position, sending $N$ to 
infinity first, before taking the limit of zero pinning field.  This is the 
usual way of identifying 
the phase transition.

In order to get around the problem of the description of the amorphous solid 
phase, a simple method 
has been developed in the spin glass context.  Pictorially, one could say that 
although we do not 
know the conjugate field, the system itself knows it.  The idea, borrowed from 
spin glass theory 
\cite{Toulouse,carparsour}, is then to consider two copies of the system, with 
an infinitesimal 
extensive attraction.  One then identifies the transition temperature from the 
fact that the two 
replicas remain close to each other in the limit of vanishing coupling (having 
sent $N$ to infinity 
first).

In the case of  glasses we can thus consider two identical
systems of particles, $\{ x_j \} $ and $\{ y_j \} $,
with a total energy function:
\be
E=\sum_{1 \le i \leq j \le N}( v(x_i-x_j) +v(y_i-y_j))+\eps \sum_{i,j} 
w(x_i-y_j)
\ee
where we have introduced a small attractive potential $w(r)$ between
the two systems. The precise shape of $w$ is irrelevant, insofar as we shall
be 
interested
in the limit $\eps \to 0$, but its range should be of order or smaller than
the 
typical 
interparticle 
distance. The order parameter is then the correlation function between the two 
systems:
\be
g_{xy}(r)=\lim_{\eps \to 0} \lim_{N \to \infty} 
{1 \over \rho N} \sum_{ij} <\delta(x_i-y_j-r)>
\ee
In the liquid phase this correlation function is identically equal to one,
while 
it
has a nontrivial structure in the glass phase, reminiscent of the pair 
correlation
of a dense liquid, but with an extra peak around $r \simeq 0$. Let us notice 
that we
expect a discontinuous jump of this order parameter at the transition, in spite
of its being second order in the thermodynamic sense. The existence of a non 
trivial order
parameter is associated with the spontaneous breaking of a symmetry: For 
$\eps=0$, with
periodic boundary conditions, the system is symmetric under a global
translation 
of
the $x$ particles with respect to the $y$ particles. This symmetry is 
spontaneously broken in 
the low
temperature phase, where the particles of each subsystem tend to sit in front of
each other. One could equally use the Fourier transform of this 
cross-correlation,
which then gives back, but in an equilibrium framework, the Edwards Anderson 
order
parameter defined in (\ref{EAvrai}).

\subsection{Thermodynamics below $T_K$: replicas}

The previous method is a reasonable definition of an equilibrium order
parameter 
which can be used 
in simulations or in analytic studies in order to identify the phase
transition 
arriving from the 
liquid phase.  However this technique can be improved in order to study 
quantitatively the glass 
phase itself.

Let us assume that in the glass phase there exists a non zero configurational 
entropy, as introduced 
above.  Clearly the knowledge of this configurational entropy as a function of 
free-energy and 
temperature, $S_c(f,T)$, will allow us to reconstruct all the interesting 
thermodynamic properties 
of the system.  It has been realised by Monasson\cite{remi} that the 
configurational entropy can be 
reconstructed from a study of an arbitrary number, $m$, of copies of the
system, 
when they are 
constrained to be in the same state.  As we will need to analytically continue 
the results in $m$, 
we shall call the copies 'replicas'.  An alternative and related method is to 
introduce a real 
coupling of the system to another system which is thermalized\cite{pot}; this 
has been used recently 
in order to study the glass phase \cite{FRAPA,CarFraPar}.  The formulation
which 
we present here is 
slightly different from, but equivalent to, that of \cite{remi}.

The basic idea is extremely simple.  Instead of two copies of the system, let
us 
consider $m$ copies 
which are constrained to stay in the same minimum.  We shall discuss below how 
one can achieve this 
constraint, but let us first discuss the physics of this constrained system.  
Its partition function 
is:
\be
Z_{m} = \int_{f_m}^{f_M} df \ e^{-N [m \beta f- S_c(f,T)]}\label{MONA}
\ee
The dependence on the number $m$ of replicas of the total free
energy,
\be
\Phi(m,T)= -{1 \over \beta N} \log Z_m \approx
\min_{f}(m \ f - T S_c(f,T)) \ ,
\ee
allows to compute  the configurational entropy $S_c(f,T) $ as a function of the
free energy, using: 
\bea
{\partial \Phi(m,T) \over \partial m}&=& f\\
{m^{2} \over T} {\partial \phi(m,T) \over \partial m}&=& S_c  \ ,
\label{conf_entr_gene}
\eea
where $\phi(m,T)$ is the free energy per particle:
\be
\phi(m,T) ={ \Phi(m,T) \over m} \ .
\ee

If the glass transition is due to the entropy crisis described in the previous 
section (and this is 
our main hypothesis), then the crucial quantity is the value of the slope 
$s_0(T)$ of the 
configurational entropy at the lowest free energy:
\be
s_0(T)\equiv{\partial S_c \over \partial f }(f_0(T))
\ee
The usual glass transition is determined by $T_K s_0(T_K)=1$. For
the replicated and constrained system, the phase transition temperature
 $T^{(m)}$ depends on the number $m$ of replicas and is
determined
by (see fig. \ref{sigma_qualit}):
\be
T^{(m)} s_0(T^{(m)})=m
\label{Tm}
\ee
It is very natural to assume that $s_0(T)$ is a smooth function of 
temperature, going to a 
constant
at zero temperature  (we shall check this hypothesis self-consistently
later). Then we see that,
%C 
when $m$ is
continued analytically to real values, {\it smaller} than unity,
one can have $T^{(m)}<T_K$. 
The replicated and constrained system can thus be in the 
liquid phase for temperatures {\it smaller} than the glass transition
temperature 
$T_K=T^{(1)}$: it is 
then
made up of molecules, each of which contains one atom of each replica,
but these molecules are in a liquid state.
 The basic reason for this
 %C
  crucial fact 
  is that for
$m<1$ the effective interaction potential (assuming for simplicity molecules
of 
very small radius)
is decreased from $v(r)$ to $m v(r)$, thus displacing the glass
transition to lower temperatures. 

\begin{figure}
\hbox{\epsfig{figure=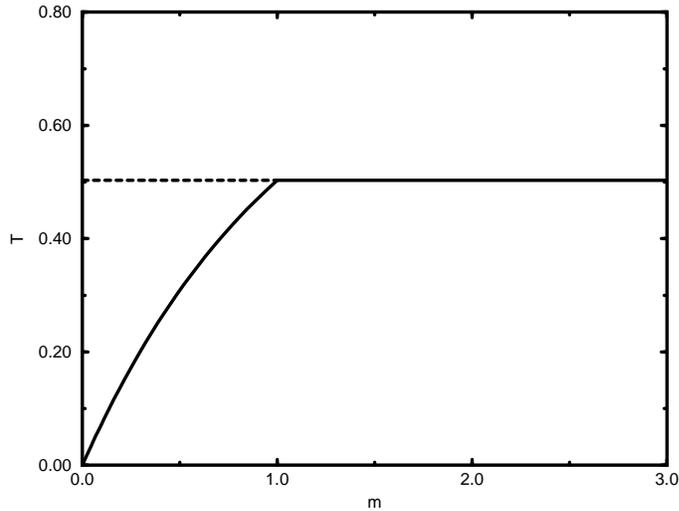,width=8cm,angle=-90}}
\caption{A sketch of the typical phase diagram in the temperature-$m$ plane,
for 
a system
with $m$ weakly coupled replicas.  In the whole high 
temperature region above 
the full line the system is in a liquid phase. There are two liquid phases, 
above the horizontal line $T=T_K$ the various replicas are not correlated in
the limit of the coupling $\eps$ going to zero. On the contrary the liquid state
at $ m<1$, in the region between the full line and the dashed line, is a 
molecular
liquid where the various replicas form molecular bound states.
The low temperature region below the full line (characterized by $m=m^*(T)$)
is the glass phase.  
In this glass phase, for a given 
temperature,  the free energy per replica 
is $m$ independent. Therefore one can deduce the free energy of the glass (
with $m<1$ and $T<T_K$) from the knowledge of the free energy in the 
molecular liquid.}
\label{fig_Tc_m} 
\end{figure}

%C Big
We are interested in the free energy in the glass phase,
therefore in the region $m=1$ and $T<T_K$. This free energy
cannot be computed from that of the liquid with $m=1$, $T>T_K$
because of the phase transition at $T_K$. However we shall now show that
one can deduce it from the free energy of the molecular fluid
at $m<1$.  
This molecular fluid with $m<1$ has a transition
to a glass state at the temperature $T=T^{(m)}<T_K$. 
Inside the glass phase,
thus for $T<T^{(m)}$, the 
free energy of the replicated and constrained system
 is given  by the condition
\be
S_c(f,T)=0 \label{ZERO}
\ee
and it is {\it independent} on $m$. 

Let us now look  at the phase diagram at
a fixed temperature $T<T_K$, varying $m$ (see fig. \ref{fig_Tc_m}).
The free energy per particle 
$\phi(m,T)$ of the molecular liquid is  an increasing function
of $m$ at small $m$, which  reaches a maximum
at a point $m^*<1$ where the glass transition takes place 
(obviously $m^*$ is the solution of: $T \; s_0(T) = m^*$). As
the free energy in the glass phase is $m$ independent, the liquid
free energy at the transition $\phi(m^{*},T)$ (which is
equal to the glass free energy at the transition)
is equal, for $T<T_{K}$, to  the free energy $\phi(m=1,T)$ 
of the glass at the temperature 
$T$. We have thus shown that the knowledge of the free energy of the
molecular liquid, $\phi(m,T)$, allows to compute the free energy of the glass.

These basic observations are at the heart of our strategy for computing 
properties of the glass 
phase.  We shall write down more explicit formulas in our case below.  
We would like first to make three comments on this approach:
\begin{itemize}
\item
For $T<T_K$ and
 $m>m^*$, the free energy $\phi(m)$ is 
constant and  
{\it  larger} than the analytic continuation of the free energy $\phi(m)$ of
the 
molecular liquid.  If one 
would have followed this molecular liquid in the region $m>m^*$, one would
have 
found that $\partial 
\phi/ \partial m <0$, predicting a negative configurational entropy.  Instead, 
the glass transition 
occurs and the configurational entropy sticks to zero in the whole glass
phase.  
The fact that the 
free energy in the glass phase is {\sl larger} than the analytic continuation 
from the high 
temperature phase explains why the specific heat has a discontinuity {\sl 
downward} when we decrease 
the temperature.  This is in variance with what happens generally in other 
transitions (at least in 
the mean field approximation) where the free energy in the low temperature
phase 
is {\sl smaller} 
than the analytic continuation from the high temperature phase and the
specific 
heat has a 
discontinuity {\sl upward} when we decrease the temperature.

\item
In practice in order to try to constrain the systems to be in the same state, 
one   introduces some small attractive coupling,
of order $\epsilon$, between the replicas. 
It is thus important to understand when this coupling 
leads to a molecular liquid.
The phase diagram shown in fig.\ref{fig_Tc_m} can be 
conjectured from the following elementary study
of the free energy, confirmed by
exact computations of mean field discontinuous spin glasses
\cite{kurparvir,pot,FRAPA,MMpspin}. There are a priori four possible cases.  
If the $m$ replicas are in 
the same state, the free energy is $\Phi=\min_{f}(m \ f - T S_c(f,T))- m 
(m-1)\eps$.  If they are 
in different states, the free energy is $\Phi=m \min_{f}( \ f - T S_c(f,T))$. 
On top of this, the free energy minimum can either stick to $f_0$ (glass
phase) or be at a value $f$ larger than $f_0$ (liquid).
One just needs to find out which situation actually minimizes the free 
energy, for given values of $m$ and $T$.  The
solution is displayed in fig.\ref{fig_Tc_m}, showing that there is an
intermediate molecular liquid phase at $m<1$.

\item 
The 'replicas' 
which we introduce here play a slightly different role compared to the ones
used 
in disordered 
systems: there is no quenched disorder here, and no need to average a
logarithm 
of the partition 
function.  `Replicas' are introduced to handle the problem of the absence of 
description of the 
amorphous state.  We do not know of any other procedure to characterize an 
amorphous solid state in 
the framework of equilibrium statistical mechanics.  There is no `zero
replica' 
limit, but there is, 
as in disordered systems, an analytic continuation in the number of replicas.  
We shall see that 
this continuation looks rather innocuous.

\end{itemize}

%C Basic hypothesis sent in discussion

\section{The replica approach to structural glasses: general formalism}
In this section we write down the formulas corresponding to the replica approach
 introduced in the previous section. We keep here to the case of 
 simple glass formers consisting of $N$ particles
interacting by a pair potential $v(r)$ in a space of dimension $d$.

\subsection{The partition function}
The usual partition function, used e.g. in the liquid phase, is
\be
Z_{1} \equiv {1 \over N!} \int \prod_{i=1}^N (d^d x_i) \  e^{-\beta H}
\ee
We wish to study the transition to the glass phase through the onset of an
off-diagonal correlation in replica space. We use $m$ replicas and  introduce
the Hamiltonian of the replicated system:
\be
H_m=\sum_{1 \le i<j \le N} \sum_{a=1}^m v(x^a_i-x^a_j) + \sum_{j_1...j_m \in
\{ 
1,...,N\} }
 W(x^{1}_{j_1},...,x^{m}_{j_m})
\ee
where $W$ is an attractive interaction. The precise form of $W$ is unimportant:
it should be a short range attraction respecting the replica permutation
symmetry, 
and its strength  which will be 
sent to zero in the end. For instance one could take
\be
W(r^1,...,r^m)= \eps \sum_{1 \le a < b \le m} w(r^a-r^b)
\ee
with $w(r)$ a smooth short range two body attraction.

The partition function of the replicated system is
\be
Z_{m} \equiv {1 \over N!^m} \int \prod_{i=1}^N \prod_{a=1}^m (d^d x^a_i) \  
e^{-\beta H_m}
\ee

The order parameter is the  generalised cross correlation:
\be
\rho(r^1,...,r^m)=  {1 \over N} 
\sum_{j_1...j_m}<\delta(x_{j_1}^1-r^1)...\delta(x_{j_m}^m-r^m)>
\ee
where the average is the Boltzmann-Gibbs average with the measure proportional 
to
$\exp(-\beta H_m)$.

\subsection{Molecular bound states}
At low enough temperature, we expect that the particles in the different 
replicas will 
stay close to each other due to the joint effect of the small inter-replica 
attraction
and the intra-replica interactions: when the system is in the glass phase, the 
role
of the attractive term $W$ will be to insure that all replicas fall into the 
same glass state,
so that the particles in different replicas stay at the same place, apart from 
some
thermal fluctuations: 
%C
A vanishingly small  interaction between replicas will give
rise to a strong correlation. As the thermal fluctuations are relatively small 
throughout
the solid phase (one can see this for instance from the Lindeman criterion), one
can identify the molecules and relabel all the particles in the various
replicas 
in such a way
that the particle $j$ in replica $a$ always  stays close to particle $j$ in 
replica $b$.
All the other relabelings are equivalent to this one, producing a global
factor 
$N!^{m-1}$
in the partition function. 

We therefore need to study a system of molecules, each of them consisting of
$m$ 
atoms
(one atom from each replica).
It is natural to write the partition function in terms
of the variables $r_i$ which describe the centers of masses of the molecules,
and the relative coordinates $ u_i^a$, with $x_i^a=r_i+u_i^a$ and $\sum_a 
u_i^a=0$:
\bea \nn
Z_{m}&=& {1 \over N!} \int \prod_{i=1}^N \(( d^dr_i \)) 
 \prod_{i=1}^N \prod_{a=1}^m
(d^d u_i^a) \prod_{i=1}^N \((m^d \delta(\sum_a u_i^a)\)) \\
& & \exp\((-\beta \sum_{i<j,a} v(r_i-r_j+u^a_i-u^a_j) - \beta
\sum_i W(u^1_i,...,u^m_i) \))
\label{Zzu}
\eea

\section{ The small cage expansion}
In order to transform these ideas into a tool for doing explicit computations
of 
the thermodynamic 
properties of a glass we have to use an explicit method for computing the free 
energy as function 
of the temperature and $m$.  As is usually the case, in the liquid phase exact 
analytic computations are not 
possible and we have to do some approximations. In this section we shall use
the 
fact that 
 the thermal fluctuations of the particles in the glass
 are small  at low enough temperature: the size of
 the `cage' seen by each particle is therefore small, allowing for a
systematic 
expansion.
 What we will be 
 describing here are the thermal fluctuations around the minimum of the 
potential
 of each particle, in the spirit of the Einstein model for vibrations of a 
crystal.

\subsection{Legendre transform}

We start from the replicated partition function $Z_{m}$ described in molecular 
coordinates in (\ref{Zzu}). Assuming that the relative coordinates $u^a_i$
are small, we can expand $W$ to leading order and write:
\bea \nn
Z_{m}&=& {1 \over N!} \int \prod_{i=1}^N \((d^dr_i\))  \prod_{i=1}^N 
\prod_{a=1}^m
(d^d u_i^a) \prod_{i=1}^N \((m^d \delta(\sum_a u_i^a)\)) \\
& & \exp\((-\beta \sum_{i<j,a} v(r_i-r_j+u^a_i-u^a_j) -
{1 \over 4 \alpha} \sum_i
\sum_{a,b} (u^a_i-u^b_i)^2 \))
\label{Zzu_exp}
\eea
In the end we are interested in the limit $(1/\alpha) \to 0$. We would like
first to define the size $A$ of the molecular
bound state, which is also a measure of the size of the cage seen by each atom 
in the glass, by:
\be
{\partial \log Z_{m} \over \partial (1/\alpha)} \equiv {m (1-m) \over 2} d  N A
=-{1 \over 4} \sum_i \sum_{a,b} \la (u_i^a-u_i^b)^2 \ra
\label{Adef}
\ee
(d is the dimension, N is the number of particles).
We Legendre transform the free energy $\phi(m,\alpha)=-(T/m) \log Z_{m}$, 
introducing
the thermodynamic potential per particle $\psi(m,A)$:
\be
 \psi(m,A)= \phi(m,\alpha)+T d  { (1-m) \over 2} {A \over \alpha}
\ee
What we want to see is whether there exists a minimum of $\psi$ at a finite 
value
of $A$.

 At low temperatures, this minimum should be at small $A$, and
so we shall seek an expansion of $\psi$ in powers of $A$. 
%C
It turns out that this can be found by an
expansion of $\phi$ in powers of $\alpha$, used as an intermediate 
bookkeeping in order to generate the low temperature expansion. 
This may look confusing since
we are eventually going to send $\alpha$ to $\infty$. However this method is
nothing but a usual low temperature expansion in the presence of an
infinitesimal breaking field. For instance if one wants
to compute the low temperature expansion of the magnetization in a 
$d$-dimensional
Ising model in an infinitesimal positive magnetic field $h$, the main point is
that the magnetisation is close to one. One can organise the expansion
by studying first the case of a large magnetic field, performing the
expansion in powers of $\exp(-2h)$, and in the end letting $h \to 0$. A little
thought shows that the intermediate -large $h$- expansion is just a
bookkeeping device to keep the leading terms in the low temperature expansion.
What we do here is exactly similar, the role of $h$ being played by $1 / 
\alpha$.

\subsection{Zeroth order term}
We use the equivalent form:
\be
Z_{m}(\alpha)= {1 \over N!} \int \prod_{i=1}^N \prod_{a=1}^m
(d^d u_i^a)
\prod_i {d^dX_i \over \sqrt{2\pi \alpha \over m^2}^d}
\exp\((-\beta \sum_{i<j, a} v(x_i^a-x_j^a)-{m \over 2 \alpha} \sum_{i,a} 
(x_i^a-X_i)^2\)) \ .
\ee
For $\alpha \to 0$, the identity 
\be
\exp\((- {m \over 2 \alpha} (x_i^a-X_i)^2\)) \simeq \(({2 \pi \alpha \over 
m}\))^{d/2}
\delta^d (x_i^a-X_i)
\ee
 gives:
\be
Z_{m}^0(\alpha) = \(({2 \pi \alpha \over m}\))^{d N m/2} 
\(( {2\pi \alpha \over m^2}\))^{-d N/2}
{1 \over N!} \int \prod_i dX_i \exp\((- \beta m \sum_{i<j} v(X_i-X_j) \)) \ .
\ee
In this expression we recognise the integral over the $X_i$'s as the partition 
function 
$Z_{liq}(T^*)$ of the liquid
at the effective temperature $T^*$, defined by
\be
 T^*\equiv T/m \ .
 \ee 
Therefore the free energy, at this leading order, can be written as:
\be
{\beta \phi^0(m,\alpha)} = {d  (1-m) \over 2 m} \log {2 \pi \alpha \over m}
-{d  
 \over 2 m}
\log(m)-{1 \over mN} \log Z_{liq}(T^*)
\ee

\subsection{First order term}
\label{small_cage}
In order to expand to next order, we start from  the representation 
(\ref{Zzu_exp})
and expand the interaction term to quadratic order in the relative coordinates:
\bea \nn
Z_{m}&=& \int \prod d^dr_i d^du_i^a \prod_i \(( m^d \delta(\sum_a u_i^a) \))
\exp\((-\beta m \sum_{i<j}
v(r_i-r_j)\))
\\ \nn
& & \exp\((-{\beta \over 2} \sum_{i<j} \sum_{a \mu \nu} (u_i^a-u_j^a)^\mu
(u_i^a-u_j^a)^\nu
\partial_\mu \partial_\nu v(r_i-r_j) 
-{1 \over 4 \alpha} \sum_{a,b} (u_i^a-u_i^b)^2\)) \ .
\label{Zzu_quad}
\eea
(The indices $\mu$ and $\nu$, running from $1$ to $d$, denote space directions).
Notice that in order to carry this step, we need to assume that the
interaction 
potential
 $v(r)$ is smooth enough, excluding hard cores.
To expand at small $\alpha$ we need the properties of the set of $m$ random
variables $u^a$ living on one site with measure $P(u) \propto \delta(\sum_a u^a)
\exp(-(1/4\alpha) \sum_{ab} (u^a-u^b)^2)$. It turns out that these are
 gaussian random variables
with a first moment which vanishes and a second moment which is equal to:
\be
\la u^a_\mu u^b_\nu \ra_0 =\(( \delta^{ab} -{1 \over m}\)) {\alpha \over m} 
\delta_{\mu \nu} \ .
\label{u_measure}
\ee
Expanding  (\ref{Zzu_quad}) to first order in $\alpha$ we have:
\be
\log Z_{m}=\log Z_{m}^0 - {\beta \over 2} \sum_{i<j} \sum_{a \mu \nu}
\la
(u_i^a-u_j^a)^\mu
(u_i^a-u_j^a)^\nu \ra_0
\la
\partial_\mu \partial_\nu v(r_i-r_j) \ra^*
\ee
where the average $\la.\ra_0$ is that for the $u$ variables with the gaussian
measure (\ref{u_measure}), and the average $\la.\ra^*$ is over the
center of mass positions $r_i$, which are those of a liquid phase thermalized at
the temperature $T^*=T/m$.

 The free energy to first order is equal to:
\be
{\beta \phi(m,\alpha)}= {d  (m-1) \over 2 m} \log{1 \over \alpha} - \alpha
\beta 
 C +
{d (1-m) \over 2 m} \log {2\pi \over m}-{d \over 2 m } \log{m} -{1 \over mN}  
\log Z_{liq}(T^*)
\ee
where the constant $C$ is proportional to the expectation value
of  the Laplacian of the potential,
in the liquid phase at the temperature $T^*$: 
\be
C \equiv  {1 \over 2} {1-m \over m^2} \sum_{j(\ne i)} \la \Delta v(z_i-z_j)\ra^*
\ee
Differentiation with respect to ${1 \over \alpha}$ gives the size of the cage:
\be
 {\beta }{\partial \phi \over \partial (1/\alpha)}= -{ (1-m) \over 2 m} d  
\alpha
+\alpha^2 \beta C=
 -{ (1-m) \over 2 } d  A
\ee
Expanding this equation in perturbation theory in $A$ we have:
\be
\alpha =m A -{ 2 \beta m^3 C \over d (m-1)} A^2
\ee
The Legendre transform is then easily expanded to first order in $A$:
\bea \nn
 {\beta \psi(m,A) }&=& {\beta \phi}(m,\alpha) + d  { (1-m) \over 2} {A \over 
\alpha}\\
&=&
{d (1-m) \over 2 m} \log(2 \pi A) - \beta m A C  +{d (1-m) \over 2 m}
-{d \over 2 m} \log m- {1 \over m N}  \log Z_{liq}(T^*)
\label{free_first_ord}
\eea

This very simple expression gives the free energy as a function of the number
of 
replicas,
$m$, and the cage size $A$. We need to study it at $m \le 1$, where we should 
maximise it
with respect to $A$ and $m$. The fact that we seek a maximum when $m<1$
instead 
of the
usual procedure of minimising the free energy is a well established fact of
the 
replica method, appearing as soon as the number of replicas is less than $1$ 
\cite{MPV}.

As a function of $A$ , the thermodynamic potential $\psi$ has a maximum
at:
\be 
A=A_{max} \equiv
{d(1-m) \over 2 \beta m^2 } {1 \over C} = {d \over \beta} {1 \over \int d^dr 
g^*(r) \Delta v(r)}
\label{A_first_ord}
\ee
where $g^*$ is the pair correlation of the liquid at the  temperature $T^*$.
A study of the potential $\psi(m,A_{max})$, which equals $\phi(m)$,
as a function of $m$ then allows to find
all the thermodynamic properties which we seek, using the formulas of the 
previous section.
 This step and the results will be explained below
in sect. \ref{results}, where we shall also compare the results to
those of other approximations.

\subsection{Higher order}
The systematic expansion of the thermodynamic potential $\psi$ in powers of
$A$ 
can be
carried out easily to higher orders. However the result involves some more
detailed properties of the liquid at the effective temperature $T^*$. For 
instance
at second order one needs to know not only the 
free energy and pair-correlation of the liquid at temperature $T^*$, but
also the three points correlation. It is certainly interesting to try to push
this expansion further, taking the information on the liquid at temperature 
$T^*$
from some numerical simulations. In this paper we have decided to stay within 
some relatively simple schemes which require only the 
knowledge of the pair-correlation $g^*(r)$. Therefore we 
shall
not pursue this higher order expansion here, leaving it for future work.

\subsection{Harmonic resummation}
 One can obtain a partial resummation of the small cage expansion described 
above
 by integrating exactly over the relative vibration modes of the molecules.
We shall use such a procedure here, which is 
a kind of harmonic expansion in the solid phase.

We work directly with $1/\alpha=0$ and
start from the replicated partition function (\ref{Zzu_quad}),
within the quadratic expansion of the interaction potential $v$ 
in the relative coordinates $u_i^a$. (Clearly it is assumed that
the $1/\alpha \to 0^+$ limit has been taken, and that its effect is
to build up molecular bound states). The exact integration over the
 gaussian relative variables gives:

\be
Z_m= {m^{Nd/2} \sqrt{2 \pi}^{N d (m-1)} \over N!} \int \prod_{i=1}^N d^dr_i
\exp\((-\beta m \sum_{i<j}
v(r_i-r_j) -{m-1 \over 2} Tr \log \((\beta M  \)) \))
\label{Z_harm_resum}
\ee
where
the matrix $M$, of dimension $Nd \times Nd$, is given by:
\be
M_{(i \mu) (j \nu)}= \delta_{ij} \sum_k v_{\mu\nu}(r_i-r_k)-v_{\mu\nu}(r_i-r_j)
\ee
and $v_{\mu\nu}(r) =\partial^2 v /\partial r_\mu \partial r_\nu$.
We have thus found an effective Hamiltonian for the centers of masses $r_i$ of 
the 
molecules, which basically looks like the original problem at the effective
 temperature $T^*=T/m$, complicated by the contribution of 
vibration modes which give the `Trace Log' term. 
We expect that this  should be a rather good approximation for the
glass phase. Unfortunately, even within this approximation,
we have not been able to compute the partition function exactly. The 
density of eigenstates of the matrix $M$ is a rather complicated object and we
have developed a simple approximation scheme in order to estimate it.

We thus proceed by using a 'quenched approximation', i.e.
 neglecting the feedback of
vibration modes onto the centers of masses. This approximation becomes
 exact close to the Kauzman temperature where $m \to 1$. The free energy is 
then:
\be
 {\beta \phi(m,T)} = -{d \over 2 m} \log(m)- { d (m-1) \over 2 m } \log(2 \pi)
-{1 \over m N} \log Z(T^*) +{m-1 \over 2 m} 
\la Tr \log \((\beta M  \)) \ra^*
\ee
which involves again the free energy and correlations of the liquid at the
temperature $T^*$.
Computing the spectrum of $M$ is an interesting problem of random matrix
theory, 
in
a subtle case where the matrix elements are correlated. Some efforts have been 
devoted
to such computations in the liquid phase where the eigenmodes are called 
instantaneous
 normal
modes \cite{INM1}. It might be possible to extend these approaches to our
case. 
Here
we shall rather propose a simple resummation scheme which should be reasonable
at high densities-low temperatures. 

%C Big
Considering first the diagonal elements of 
$M$,
we notice that in this high density regime there are many neighbours to each 
point, and
thus a good approximation is to neglect the fluctuations of these diagonal
terms and substitute them by their average value. We thus write:
\be
\sum_k v_{\mu\nu}(r_i-r_k)\simeq \delta_{\mu\nu}
{1 \over d} \int d^dr g^*(r) \Delta v(r) \equiv r_0
\label{diag_non_flu}
\ee
Here and in what follows,
we have not written explicitly the
density. We choose to work with density unity in order to
 simplify the formulae: this 
value can always be 
obtained by using an appropriate scale of length.
In the approximation (\ref{diag_non_flu}) the diagonal matrix 
elements are all equal and can be factorized, leading to:
\be
\la Tr \log \((\beta M  \)) \ra^* =
N d \log (\beta   r_0 )+
\la Tr \log \(( \delta_{ij} \delta_{\mu\nu} - {1 \over r_0}
v_{\mu\nu}(r_i-r_k) 
\)) \ra^*
\label{tracelog}
\ee
This form lends itself to a perturbative expansion in powers of
$1/r_0$. The computation of the $p$-th order term in this expansion, 
\be
{\cal T}_p \equiv {(-1)^{p-1} \ over p r_0^p} \la 
\sum_{{i_1...i_p} \atop {\mu_1...\mu_p}} 
v_{\mu_1 \mu_2}(r_{i_1}-r_{i_2})...
v_{\mu_{p-1} \mu_p}(r_{i_{p-1}}-r_{i_p})
v_{\mu_p \mu_1}(r_{i_p}-r_{i_1})
\ee
still involves the $p$-th order
correlation functions
of the liquid at $T^*$. We have approximated this full correlation
by introducing a  simple `chain' approximation involving only the pair 
correlation.  This chain approximation consists in replacing,
for $p>2$,
the full correlation by a product of pair correlations.
It selects  those contributions which survive in the high 
density limit; systematic corrections could probably be computed in the 
framework of the approach 
of 
\cite{Stratt}, we leave this for future work. 
 Within the chain approximation, ${\cal T}_p$ is approximated by:
\bea
{\cal T}_p &=& \sum_{\mu_1...\mu_p}
\int dx_1...dx_p \  g^*(x_1,....,x_p) \  [v_{\mu_1 \mu_2}(x_1-x_2)...
v_{\mu_{p-1} \mu_p}(x_{p-1}-x_p )v_{\mu_p \mu_1}(x_p-x_1)]
\\ 
&&\simeq
\sum_{\mu_1...\mu_p}
\int dx_1...dx_p \  [g^*(x_1-x_2) v_{\mu_1 \mu_2}(x_1-x_2)]...[g^*(x_p-x_1) 
v_{\mu_p \mu_1}(x_p-x_1)] \ .
\eea
In this last form we need to compute a convolution which can be factorised
through the introduction of the Fourier transform of the pair correlation
function. We thus introduce 
the Fourier transformed functions $a$ and $b$ which are defined
from the pair correlation $g^*(r)$ by:
\be
\int d^d r \  g^*(r) v_{\mu\nu}(r) e^{ikr} \equiv \delta_{\mu\nu} \ a(k) +
\(( {k_\mu k_\nu \over k^2} -{1 \over d} \delta_{\mu \nu}\)) b(k) \ .
\label{defab}
\ee
In terms of these Fourier transforms, the $p$-th order term in the $1/r_0$
expansion is simply
\be
{\cal T}_p=
\intk \((a(k)+{d-1 \over d} b(k)\))^p + (d-1) \intk  \((a(k)-{1 \over d} \ .
b(k)\))^p \ ,
\ee
and the summation of the series over $p$ is easily done, so that
the free energy per particle within the chain approximation of the harmonic
resummation  is:
\bea
 {\beta \phi(m,T)} &=& -{d \over 2 m} \log(m)- { d (m-1) \over 2 m } \log(2 \pi)
-{1 \over m N} \log Z(T^*) + {d (m-1) \over 2 m } \log (\beta r_0)
\\
\nn
&+&
{(m-1) \over 2 m } \intk  \(( L_3 \(({a(k)+{d-1 \over d} b(k) \over r_0}\))
+
{(d-1)} L_3 \(({a(k)-{1 \over d} b(k) \over r_0}\)) \)) \\ 
&-&
{(m-1) \over 4 m } \int d^dr g(r) \sum_{\mu\nu} {v_{\mu\nu}(r)^2 \over r_0^2}
\label{chain}
\eea
where the function $L_3$ is defined as:
\be
L_3(x)=\log(1-x)+x+{x^2/2}
\ee
We can thus compute the replicated free energy $F_m$ only from the knowledge
of the free energy and the pair correlation of the liquid
at the effective temperature $T^*$.
The results will be discussed in section \ref{results}.

\section{A systematic Approach: molecular HNC closure}
\label{HNC}

\subsection{Density functional}
As we have seen before, one can choose as an order parameter the generalised
inter-replica correlation, 
deduced from the original partition function by the functional derivative:
\be
\rho(r^1,...,r^m) =-{1 \over \beta} {\delta \log Z_m \over \delta 
W(r^1,...,r^m)}
\ee

In order to study the free energy at fixed order parameter,
one can perform the functional Legendre transform:
\be
{ \psi}[\rho]=-{T \over m} \log Z_{m}[\phi] -
{1 \over m} \int \ dr^1...dr^m  \rho(r^1,...,r^m) W(r^1,...,r^m)
\ee
and the aim is to optimize this new function with respect to $\rho$.

In the ideal case where there are no interactions, this 
  thermodynamic potential is:
\be
\psi^{id}[\rho]= {T \over m} \int dr^1...dr^m \rho(r^1,...,r^m) \log
{\rho(r^1,...,r^m) \over e}
\ee
We need to add to this piece the part which comes from the interactions.
This is non trivial; in the next section we shall use the HNC approximation
for this function.

\subsection{Molecular HNC equations}
The free energy in the HNC approximation is derived in the appendix I. It
is a functional of the molecular density $\rho(x)$ and the two point
correlation $g(x,y)\equiv1+h(x,y)$. Here and in the following, the letters
$x$, $y$ and $z$  without any index denote $m d -$ dimensional vectors
(e.g.: $x=x^1,...,x^m$). The molecular density is our order parameter. The 
result for $\psi$
is:
\bea \nn
{\beta \psi } &= &{1 \over 2 m} \int dx dy \rho(x) \rho(y)
\[[ g(x,y) \log g(x,y)-g(x,y)+1+\beta v(x,y) g(x,y)\]]\\ 
&-&{1 \over 2 m} Tr \(( \log(1+h \rho)-h \rho+{1 \over 2} h \rho h \rho \)) 
+{ 1 \over m} \int dx \rho(x) \log { \rho(x) \over e}
\label{Ghnc}
\eea
where the potential
is $v(x,y)=\sum_a v(x_a-y_a)$. In the trace term all products are
convolutions. 
For
instance the lowest order term in the small $\rho$ expansion of the trace is:
\be
-{1 \over 3} \int d^{md}x \  d^{md}y \  d^{md}z \ 
h(x,y)\rho(y) h(y,z) \rho(z) h(z,x) \rho(x)
\ee

We would like to optimize the thermodynamic potential $\psi$ with respect to
the 
molecular
density $\rho(x)$ and the two point function $g(x,y)$. We shall work at low
temperatures for which $\rho$ should be nearly gaussian.
We thus choose an Ansatz for $\rho$ of the type (always with a choice of
average density equal to one):
\bea
\rho(x)&=& \int d^dX \prod_{a=1}^m \(({\exp\((-(x^a-X)^2/(2 A)\)) \over 
\sqrt{2 \pi A}^d} \)) \\
&=& 
\(({ 2 \pi A \over m}\))^{d/2}
\(({ 2 \pi A }\))^{-md/2}
\exp\((-{1 \over 4 A m} \sum_{ab} (x^a-x^b)^2\))
\label{density}
\eea
where the molecular density is parametrized by the single parameter $A$.

The ideal gas contribution (last term in (\ref{Ghnc}) gives:
\be
\int \prod_a d^d x^a \rho(x) \log {\rho(x) \over e}= N \(( {d \over 2} (1-m) 
\log(2 \pi A)
+{d\over 2} (1-m) -{d\over 2} \log m -1 \))
\ee

The interaction term is more complicated, and we have only succeeded in 
optimising
it in the small cage regime.

\subsection{Second order small cage expansion}
Here we  shall solve in general for $g$
in the limit of small  cage radius $A$, expanding in powers of $A$. 

As usual we go to molecular coordinates, introducing $x^a=X+u^a$ and 
$y^a=Y+v^a$,
with the constraints: $\sum_a u^a= \sum_a v^a=0$. The molecular density 
(\ref{density}) depends only on the relative coordinates:
\be
 \rho(u) \equiv \rho_0 m^d  \delta(\sum_a u^a) \(({2 \pi A \over m}\))^{d/2}
(2 \pi A)^{-dm/2} \exp\((-{1 \over 4 A m} \sum_{ab} (u^a-u^b)^2\))
\label{rhodeu}
\ee
The $u$'s are thus gaussian distributed with a second moment:
\be
\la u_a^\mu u_b^\nu \ra = A \((\delta_{ab}-{1\over m}\)) \delta_{\mu\nu}
\label{u2av}
\ee

We shall expand the two point correlation in powers of the relative coordinates,
using the notations:
\bea
g( \{ X+u^a \} ,\{ Y+v^a \} )&=& G(X-Y)+ \sum_{\mu\nu}
S_{\mu\nu}(X-Y) \(( \sum_a [u^a_\mu u^a_\nu+v^a_\mu v^a_\nu]-2 K_{\mu\nu}\))
\\
&+& \sum_{\mu\nu}
T_{\mu\nu}(X-Y) \(( \sum_a [(u^a_\mu-v^a_\mu) (u^a_\nu- v^a_\nu)]-2 
K_{\mu\nu}\))
\label{STdef}
\eea
where  the constant
$K_{\mu\nu}$ is chosen in such a way that, for any $A$:
\be
\int du \rho(u) \int dv \rho(v) \ g(X+u^1,...,X+u^m;Y+v^1,...,Y+v^m)= G(X-Y)
\label{constraints}
\ee
The constant turns out to be: 
\be
K_{\mu\nu}= A (m-1) \delta_{\mu\nu} \ .
\label{Kvalue}
\ee

It is not difficult to see that, thanks to the constraint (\ref{constraints}),
the knowledge of the functions $S$ and $T$ is enough to compute the free
energy 
to order $A^2$.
This computation is done in the
appendix II. Here we just give the result.
We write the free energy to second order in the form:
\be
{\beta \psi} = \beta F_0 +\beta F_0' +\beta F_1 +\beta F_2
\label{def_free_ord2}
\ee

The zeroth order terms are:
\be
\beta F_0={d \over 2} {1-m \over m} \log(2 \pi A)
+{ (d-2) \over 2} {1-m \over m} -{d \over 2 m} \log m
\ee
\bea \nn 
\beta F_0'&=& {1 \over 2 m} \int {d^dk \over (2 \pi)^3} \(( -\log(1+H(k))+H(k)
-H(k)^2/2 \))\\ 
&+& {1 \over 2 m} \int d^dr \((G(r) \log G(r) -G(r)+1+\beta m v(r) G(r)\))
\label{hnc0}
\eea
where $H(r)\equiv G(r)-1$, and $H(k)$ is the Fourier transform of $H(r)$.
It is clear from(\ref{hnc0}) 
that the zeroth order correlation function $G(r)$ is exactly the
pair correlation of the liquid at the effective temperature $T^*=T/m$ in the
HNC 
approximation, 
we thus recover our previous results.

The first order correction is:
\be
\beta F_1= \beta A {m-1 \over 2 m} \int d^dr G(r)  \sum_\mu v_{\mu\mu}(r)
\label{hnc1}
\ee
At this order we can easily optimize the free energy with respect to $G(r)$,
and with respect to the cage size $A$. We get back the same result for
$A$ and the free energy as we had in the direct first order small cage expansion
 (\ref{A_first_ord}).

The advantage of this molecular HNC approach is that we can compute the second 
order
term without needing to solve for three point correlations in the liquid.
The second order correction is:
\bea 
\beta F_2 &=& A^2 {m-1 \over  m} \int d^dr {1 \over G(r)} \sum_{\mu\nu}
\(( S_{\mu\nu}(r)^2 +2 S_{\mu\nu}(r) T_{\mu\nu}(r)+ 2 T_{\mu\nu}(r)^2\))
\\ \nn
&+&  A^2 {m-1 \over  m} \int d^dr \sum_{\mu\nu}
\(( S_{\mu\nu}(r)+ 2 T_{\mu\nu}(r)\)) \beta v_{\mu\nu}(r)
\\ \nn
&+&  A^2 {(m-1)^2 \over 4 m^2} \int d^dr G(r) \sum_{\mu\nu} \beta 
v_{\mu\mu\nu\nu}(r)
\\ 
&-&  A^2 {m-1 \over  m} \intk \sum_{\mu\nu} \((S_{\mu\nu}(k)+T_{\mu\nu}(k)\))^2
{H(k) \over 1+H(k)}
\label{hnc2}
\eea
The stationarity conditions on $S$ and $T$ are easily solved. One finds:
\be
T_{\mu\nu}=-{1 \over 2} G(r) \beta v_{\mu\nu}(r)
\label{eq_T}
\ee
while $S+T$ is the solution of the linear equation:
\be
{S_{\mu\nu}+T_{\mu\nu} \over G} +{1 \over 2} \beta v_{\mu\nu}=
\intk e^{ikr} \((S_{\mu\nu}(k)+T_{\mu\nu}(k)\)) {H(k) \over 1+H(k)}
\label{eq_ST}
\ee
The equation for $G$ is also easily found. Expanding $G=G_0+A G_1$,
one sees that $G_0$ is the  pair correlation $g^*$ of the liquid at
 temperature $T/m$, while the correction $G_1$ is the solution of the linear 
equation:
\be
{G_1(r) \over G_0(r)} + \beta (m-1) \sum_{\mu} v_{\mu\mu}(r)=
\intk e^{ikr} {H_0(k)(2+H_0(k)) \over (1+H_0(k))^2} G_1(k)
\label{eq_G1}
\ee

The solution of these equations and the physical consequences are discussed in 
the next
section.

\section {Results}
\label{results}
In this section we indicate how to obtain the thermodynamic properties of
the glass within each of the previous approximation scheme, and we give the
results. 

\subsection{Methodology}

We have developed in this paper three approximation schemes. 

The  small cage expansion has
been carried out directly to first order in section \ref{small_cage}, and
agree 
with the
first order expansion within the molecular HNC approach. Within this first order
approximation, the cage size is given explicitly in (\ref{A_first_ord}) and 
the corresponding free energy $\phi(m)$ is given in (\ref{free_first_ord}). We 
need to study
the $m$ dependance of $\phi$. Clearly the only ingredients we need are the
free 
energy
and pair correlations of the liquid at the temperature $T^*=T/m$, which is a 
temperature
which lies in the range of the glass transition temperature, as we shall see. 
These
properties of the liquid could be obtained by various means; here we have used 
the HNC
closure for the pair correlation and the corresponding free energy in order to 
get them.
(obviously one could try to use better schemes of 
approximation
for the liquid, depending on the form of $v(r)$, in order to improve the
results; our point here is not to try to get the most precise results, but
 to show the feasibility of a quantitative  computation of glass properties
using the simplest approximations).
Given the temperature $T$, the procedure is the following: we vary the value
of 
$m$, and
for each value we can compute the cage size $A$ and the free energy $\phi(m)$. 
As expected on
general grounds (see section \ref{strategy}), we find a free energy which 
increases with $m$
until it reaches the critical value $m^*(T)$ (such that (\ref{Tm}) holds),
which 
is the phase
transition boundary. This critical value is defined by $\partial \phi
/\partial 
m=0$.
The configurational entropy is given by the solution of the two general 
equations
(\ref{conf_entr_gene}), and the free energy of the glass is nothing but 
$\phi(m^*,T)$.
We get the internal energy and specific heat by differentiating the free
energy. 
The critical (Kauzman) temperature $T_K$ is defined by $m^*(T_K)=1$. 

The second approximation scheme is the harmonic resummation method. Again we 
have
an explicit form (\ref{chain}) for the free energy 
per particle $\phi(m)$ only from the knowledge of the free energy and the pair 
correlation
of the liquid at $T^*$. Having this $m$ dependance the procedure to get the 
thermodynamic
results is entirely the same as that of the first order result.

The third approximation scheme is obtained by the expansion of the molecular
HNC free energy to second order in the cage size, as described in section 
\ref{HNC}.
For  given values of the temperature $T$ and the number of replicas $m$, we 
first
solve the standard HNC equations giving the pair correlation $G(r)=g^*(r)$ at 
the temperature
$T^*=T/m$. Then we can compute the functions $S,T$ and the correction to the 
correlation
$G_1$ by solving the set of linear equations 
(\ref{eq_T},\ref{eq_ST},\ref{eq_G1}). The 
free energy is then  computed to second order as in (\ref{def_free_ord2}).

%C
We use the results of the second order term in the expansion in a perturbative 
way which we shall now describe. One might be tempted to use the 
free energy computed to order $A^2$ without expanding the solution to order
$A^2$. However this procedure
 is not useful because the equations truncated at the 
order 
$A^{2}$ do not have a 
solution. One must do the computation fully
perturbatively in a consistent way,
which we now explain.
 Let us define
the various terms in this free energy as
\be
 {\beta \psi(A,m)}\equiv \gamma_0 + A \gamma_ 1+A^2 \gamma_2 + \gamma_3 \log A
\ee
where the $\gamma$'s are functions of $m$ that we can compute. 
We suppose
that the $\gamma_2$ term is small and write the value $A_{max}$
which maximises
%C
\footnote{One must maximise the free energy with respect to $A$, instead
of the usual minimization procedure, whenever $m$ is less than $1$. This 
is a usual aspect of the replica method, which is here a consequence
of the fact that the free energy is proportional to $m-1$}
 the free energy as: 
\be
A_{max}=-{\gamma_3 \over \gamma_1} - 2 {\gamma_2 \gamma_3^2 \over \gamma_1^3}
\ee
giving a free energy on this maximum approximately equal to
\be
{\psi(A_{max},m)}
=\psi_1(m)+ \psi_2(m)
\ee
with
\bea \nn
\psi_1(m)&=&\gamma_0-\gamma_3 +\gamma_3 \log(-\gamma_3/\gamma1)\\
\psi_2(m)&=&\gamma_2 \gamma_3^2/\gamma_1^2
\eea 
where $\psi_2$ is the correction term.
This is a function of $m$ which we maximise in order
to find the critical value $m^*$. Writing $m^*=m_1+  m_2$, where $m_1$ is the
critical value computed to first order and $m_2$ is the correction, these
numbers satisfy the equations:
\bea \nn
0 &=&{\partial \psi_1 \over \partial m}(m_1) \\
m_2&=&- {\partial \psi_2 \over \partial m}(m_1) \(( {\partial^2 \psi_1 \over 
\partial 
m^2}(m_1)\))^{-1}
\label{mexp}
\eea
 For
consistency of this perturbative expansion,
one should then compute the saddle point value of $A$ as:
\be
A= -{\gamma_3(m_1) \over \gamma_1(m_1)} - 2  {\gamma_2(m_1) \gamma_3(m_1)^2
\over \gamma_1(m_1)^3} -m_2 {\partial \over \partial m_1}{\gamma_3(m_1) \over 
\gamma_1(m_1)}
\ee
and  the free energy of the glass as:
\be
{\psi} = \psi_1(m_1) + \psi_2(m_1)
\ee
Having the free energy as a function of $m$ we proceed as before by maximising 
it,
following exactly the same steps as for the first order computation.

\subsection{Numerical procedure}

We have studied the case of soft spheres in three dimensions interacting
through a potential $v(r)=1/r^{12}$. We work for instance at unit density, since
the only relevant parameter is the usual combination $\Gamma=\rho T^{-1/4}$.

For each of the three approximation schemes
mentioned above, we need to compute the free energy and the pair correlation of
the liquid in a temperature range close to the glass transition. We have used
the HNC approximation to get both $g(r)$ and the free energy. We have
solved the HNC closure equations numerically.
  For spherically symmetric functions in dimension three 
we use the Fourier transform for the radial dependance, in the following form:
\be
q {\bf h}(q)   = 2 \pi  \int_0^\infty dr \sin(qr) r h(r).
\ee

We discretize this formula introducing in $r$ space a cutoff $R$ and a mesh
size 
$a$.  In this way 
we have a simple formula for the inverse Fourier transform and we can also use 
the fast Fourier 
transform algorithm.  In most of the computations we have taken $a=1/32.5$ and 
$L=128*a \approx 4$.  
We have checked that dividing $a$ by 2 and multiplying $L$ by two (thus going
up 
to 512 points) does 
not alter the results.  The solution of the equations can be found either by 
using a library 
minimization program , or a program which solves non linear equations.  We
have 
found first the 
solution at low enough density and then followed it by continuity while 
gradually increasing the 
density.

The second order expansion of the molecular HNC theory requires some more
work, because we need to compute the various tensors $S_{\mu\nu}$, $T_{\mu\nu}$,
and the correction to G. After decomposing the tensors in their various
irreducible components, using rotation invariance, these components are 
discretized
on the same grid as $g(r)$ and the linear equations are solved by a standard 
library
routine.

\subsection{Critical temperature and effective temperature}
\begin{figure}
\centerline{
\hbox{\epsfig{figure=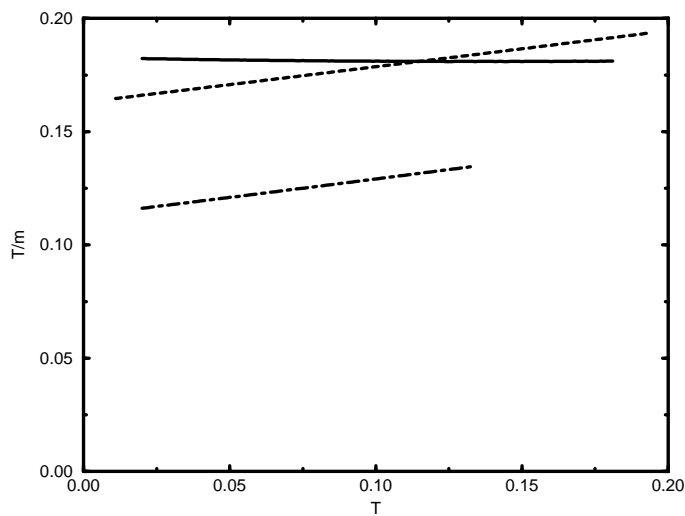,width=8cm,angle=-90}}
}
\caption{The  effective temperature of the molecular liquid at the transition, 
$T^*= T/m^*$,
 versus the temperature $T$, computed in an expansion to first 
order (dashed-dotted line) and second order(full line) in the cage size $A$,
and in the harmonic resummation (dashed line). }
\label{fig_m}
\end{figure}

We  plot in fig. \ref{fig_m} the inverse of the effective temperature $T^*$,
equal to $m^*/T$, 
versus the temperature $T$ of the thermostat. The transition temperature is 
given
by $T^*=T$. This gives the ideal glass transition temperature.
 Within the first order expansion we find $T_K \simeq .14$; the harmonic
 resummation gives $T_K \simeq .19$ and the second order perturbation theory
 is $T_K \simeq .18$  
  We see that the two best methods, the second order
 and harmonic resummation, are in good agreement and give a critical value 
 of $\Gamma$ around $\Gamma \simeq 1.52$. This value of $\Gamma$ is in good
 agreement with the published values of the glass transition of the soft
sphere 
system,
 which range around 1.6 \cite{Han}.

We also notice that the effective temperature stays relatively
constant when the 
actual 
temperature varies.
Our results are not so far from a situation in which one would have $T^*
\simeq 
T_K$,
independently from the value of the temperature $T$, which means that 
$m \simeq T/T_K$. A nearly linear variation of $m$ versus $T$ is
 often found in discontinuous spin glasses, where it
is characteristic of a free energy landscape which is totally
frozen in the whole low temperature phase \cite{GrossMez}. It is
worth noticing that such a relation has also been found for the
temperature dependance of the fluctuation dissipation ratio (although,
as this ratio is a dynamical quantity, it rather equals $T/T_D$, where
$T_D$ is the dynamical (mode-coupling) transition temperature).

\subsection{Cage size}
\begin{figure}
\centerline{
\hbox{\epsfig{figure=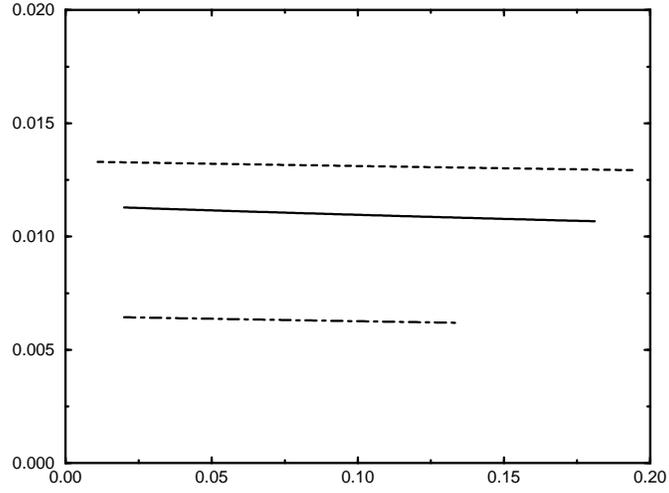,width=8cm,angle=-90}}
}
\caption{The parameter A/T versus the temperature, 
computed in an expansion to first 
order (dashed-dotted line) and second order(full line) in the cage size $A$,
and in the harmonic resummation (dashed line)
}
\label{figA}
\end{figure}

In replica space the cage size characterizes the size of the molecular bound 
state,
in the approximation of quadratic fluctuations, 
as defined in (\ref{Adef}).  Its physical meaning is easily established: In
the glass phase at low temperatures one can approximate the movement
of each atom as some vibrations in a harmonic potential
in the neighborhood of a local minimum of the energy. The typical square size
of the displacement is given by:
\be
A= \la (r_i- \la r_i \ra)^2\ra
\ee
which is the physical definition of the square size.
The cage size is plotted versus temperature in fig. \ref{figA}.
The cage size is nearly linear in temperature, as it would be in a 
$T$-independent
quadratic confining potential. This  indicates that the local confining
potential  has  little dependance on the temperature in the whole low
temperature phase.

\subsection{Free energy and specific heat}
\begin{figure}
\centerline{
\hbox{\epsfig{figure=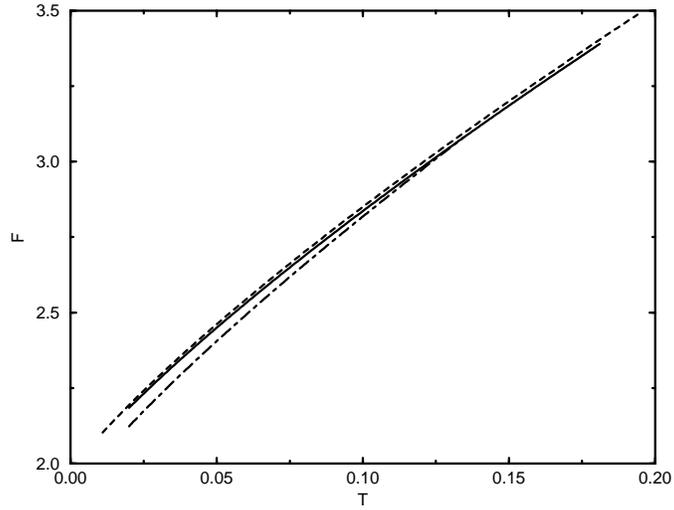,width=8cm,angle=-90}}
}
\caption{The free energy versus the temperature,
computed in an expansion to first 
order (dashed-dotted line) and second order(full line) in the cage size $A$,
and in the harmonic resummation (dashed line). }
\label{fig_free}
\end{figure}

In fig. \ref{fig_free} we plot the free energy  versus
the temperature for each of our three approximations. The strong consistency
of 
the
second order small cage expansion and the harmonic resummation are clearly seen.
The data extrapolates at zero temperature to a ground state energy of order
1.95. This is related to the typical energy of the amorphous 
packings of soft spheres.
More precisely, if we consider all the amorphous packings of soft spheres
at unit density, we can count them through the zero temperature configurational
entropy. The lowest energy at which one can find an exponentially
large number of such packings is the ground state energy of the glass state
which we find within our approximations equal to 1.95. This could be amenable
to some numerical test \cite{dald,doye,APRV}. However in order to do such a
test 
one must
remember that we have not taken into account the existence of a 
crystal: therefore  one must first
 remove all crystal like solutions, i.e.  solutions which correspond 
to a crystal with 
some local defects. These solutions can be characterized by the presence of 
delta functions at the 
appropriate values of the momenta.  
This procedure of 
identifying crystal like solutions has been explicitly done numerically in 
\cite{APRV}. Generalizing the present result to hard spheres would allow for a 
computation of random close packing density, a notion which is often
used in granular materials \cite{nowak}.

\begin{figure}
\centerline{
\hbox{\epsfig{figure=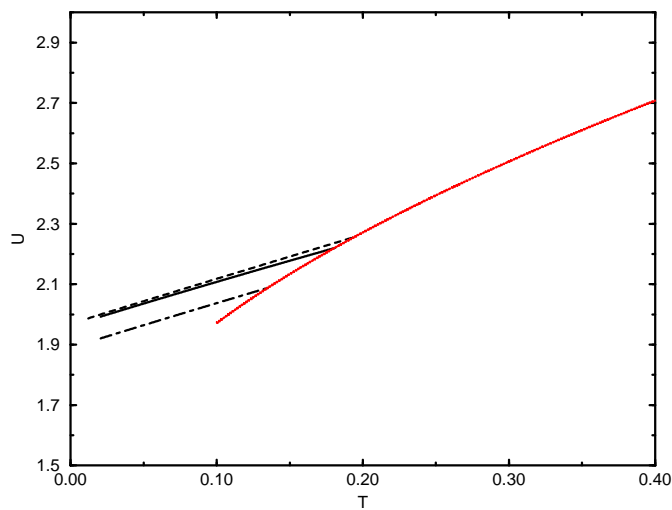,width=8cm,angle=-90}}
}
\caption{The internal energy versus the temperature,
computed in an expansion to first 
order (dashed-dotted line) and second order(full line) in the cage size $A$,
and in the harmonic resummation (dashed line). Also shown is the internal
energy 
of
the liquid (dotted line). }
\label{int_ene}
\end{figure}

In fig.\ref{int_ene} we plot the internal energy of the glass versus 
temperature,
computed in each of our approximation schemes. Also shown is the internal
energy 
of
the liquid. The internal energy is continuous at the transition.

\begin{figure}
\centerline{
\hbox{\epsfig{figure=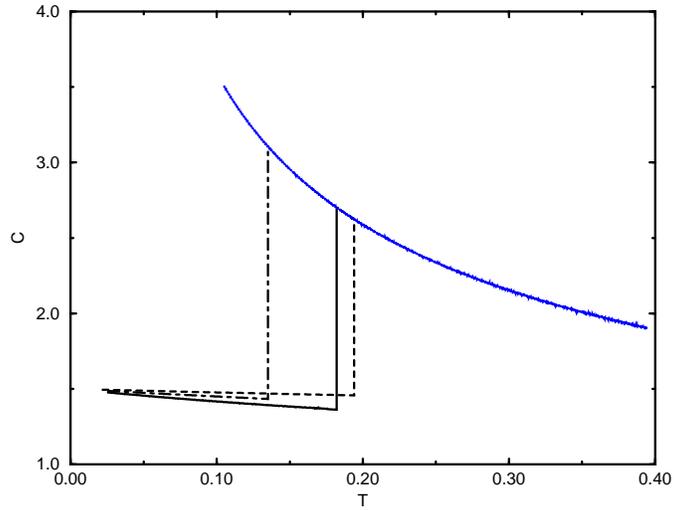,width=8cm,angle=-90}}
}
\caption{The  specific heat of the glass versus the temperature, 
computed in an expansion to first 
order (dashed-dotted line) and second order(full line) in the cage size $A$,
and in the harmonic resummation (dashed line). The dotted line is the specific
heat of the liquid.}
\label{fig_C}
\end{figure}

In fig. \ref{fig_C} we plot  the specific heat versus temperature. It is 
basically constant
and equal to 3/2. The fluctuations are numerical errors due to the
extraction of the specific heat through the 
numerical second derivative of the free energy. A specific heat
$C=3/2$ is nothing 
but the
Dulong-Petit law (we have not included the kinetic energy
 of the particles,
which would give an extra contribution of $3/2$). This result is very welcome:
in fact if we had treated the crystal at the same level of approximation as
we considered here for the glass, we would get the Einstein model for which
the specific heat is also given by the Dulong-Petit law. Thus we have found
that the specific heat of the glass is equal to that of the crystal, which is
a good approximation of the existing data. Notice that it was not
obvious at all a priori that we would be able to get such a result form our
computations, since we are performing some computations purely in the liquid
phase, with a liquid pair-correlation etc... The fact of finding the 
Dulong-Petit law is an indication that our whole scheme of 
computation
gives reasonable results for a solid phase. 
At a later stage we would like to go beyond the Dulong-Petit law and get
a better computation of the spectrum of soft vibration modes in order to get
a Debye-like law.  This is left for future work.

\subsection{Configurational entropy}
\begin{figure}
\centerline{\hbox{
\epsfig{figure=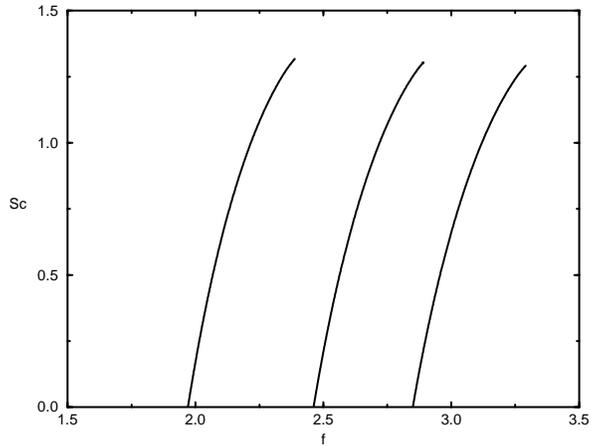,width=7cm,angle=-90}
}}
\caption{The configurational entropy $\Sigma(f)$ versus the free energy,
computed within the harmonic resummation, 
at temperatures
$T=0.,.05,.1$ (from left to right). }
\label{fig_Sc}
\end{figure}

In fig. \ref{fig_Sc} we show the configurational entropy versus the free energy
at various temperatures, including the zero temperature case. We have included
here for simplicity only the result from the harmonic resummation procedure.

We notice that the various curves corresponding to different
temperatures are not far from being just shifted one from another
by adding a constant to the free energy. This indicates that the main effect
of 
temperature 
 amounts to an additive constant in the energies of all amorphous packings. This
would be the case if the states at finite temperature could be deduced
continuously from the zero temperature amorphous packings, with an extra 
contribution
to the free energy coming from the vibrations, if the vibration spectrum is
more 
or less
state independent.

\subsection{Dynamical transition}
As we discussed in the introduction, at the mean field level  there exists a 
dynamical transition at 
a temperature $T_D$ larger than the thermodynamic transition temperature
$T_K$.  
This phase is 
characterized by the dynamic statement that a system will remain forever in
the 
same valley, and its 
free energy is greater than the equilibrium one because it misses the 
contribution of the 
configurational entropy.  It is thus evident that this dynamic phase is just a 
mean field concept, 
which should disappear when corrections, such as activated processes, due to
the 
short range nature 
of the potential, are taken into account.  However if the barriers are 
sufficiently high, metastable 
states have a very large life time and they strongly affect the dynamics.  It 
would be thus 
interesting to try to compute the `dynamic transition temperature' in these 
systems.

In the framework of the harmonic resummation one finds that the approximation 
breaks down at small 
but positive $\eps$ if the matrix of second derivatives has negative 
eigenvalues.  From this point 
of view the appearance of negative eigenvalues signal the dynamic transition.  
Unfortunately in our 
chain approximation all the eigenvalues are positive at all temperatures and
no 
dynamic phase 
transition can be seen: the free energy is always well defined for small
$\eps$. 
 This negative 
result is due to the fact that the chain approximation we use may be
reasonable 
at low temperature 
but it is certainly not good at high temperatures.  This problem will
disappear 
if one uses a better 
method to compute the spectrum, giving reasonable results also at higher 
temperatures.  On the other 
hand in the framework of the small cage expansion the perturbative method 
assumes that there is 
always a bound state.  Although this should not be true at high temperature,
the 
breakdown of this 
assumption cannot be seen in a perturbative approach.

It is clear that a study of the dynamical phase transition should be done
using 
some different tools 
than the one we have developed here.  This is not surprising: the dynamical 
phase transition is 
present at a temperature higher than the static one and the approximations
which 
we have been using 
are low temperature ones.

\section{Discussion and perspectives}
%C
Deducing the thermodynamic properties of the glass from those of a liquid
may look crazy. Of course the main trick is that we use
a molecular liquid, with a variable number $m$ of atoms per molecule,
which will have a glass transition at a temperature lower than $T_K$ 
whenever $m<1$.
We wish to underline again what is the basic hypothesis of our approach.  We 
assume that there exists a thermodynamic glass transition, which is of the
general type described in our `basic scenario'. This 
assumption means that there 
exists a path in the $m \ , \ T $ space which connects the points $m<1 \ , \ 
T^{(m)}$ to the high 
temperature region without crossing any transition.  If this is true (and
this is known to happen in many models) the situation is rather simple and 
corresponds to what is called 
in the literature {\sl one step replica symmetry breaking}.  This situation 
corresponds to the case 
in which the deep minima of the free energy are completely uncorrelated 
\cite{MPV,BouMez_extr}.  One 
could think of checking this hypothesis numerically by computing for small 
systems all the 
metastable states at zero temperature, and studying the distribution of their 
energies.
 Let us mention for 
completeness that there exist 
models in which the deep minima of the free energy are partially correlated 
(this is very probably 
the case of spin glasses \cite{FRARIE,MPRR}).  In such a case any path in the
$m 
- T $ space 
which connects the 
point $(m ,
T^{(m)} )$ to the high temperature region crosses a phase transition, and one 
would need to 
introduce a more complex construction in order to avoid this singularity.

The approach described in this paper opens the way to the computation of the 
thermodynamic properties 
of glasses at all temperatures using the generalization of the standard tools
of 
liquid theory.
%GP
Although it is not explicitly discussed in this paper, this approach allows also
the computation of the density  correlation function $g(r)$ in the glassy
phase; 
we plan to address this point in the next future.

It is clear that the results presented here  just use the simplest possible
non 
trivial approximations. Nevertheless, within these simple
approximations, we have shown that a reasonable value of the 
Kauzman temperature can be derived, as well as several thermodynamic
properties of the glass phase: the internal energy, free energy, configurational
entropy and specific heat, and the cage radius. Obviously our study so far
has been restricted to equilibrium properties, and the equilibrium
situation
is very difficult to reach experimentally. However one can think of measuring
each 
of the above properties in  numerical simulations, where the joint
use of smart algorithms and small enough system can allow to thermalize.
The extension of the present methods to binary mixtures is a work
that must be done in order to allow for a more precise comparison with
the results of numerical simulations. Some steps have 
already been done in this direction \cite{BPGM}. 

This equilibrium study is to be considered as a first step before dealing
with the  out of equilibrium dynamics. Beside the dynamics
in the low temperature phase, 
 a very interesting and open problem is the computation of the time 
dependent correlation 
functions (and as a by-product the viscosity) in the region above $T_{K}$.
However a better understanding of activated processes in this 
framework is a 
crucial prerequisite.

Within the equilibrium framework, we have implemented so far our
general strategy using rather crude methods. These should be improved,
which means that one must perform a more careful study of the molecular liquid.
  There are many directions in which one could move:
\begin{itemize}
\item Improve  the computation of the spectrum in the harmonic approximation.
This harmonic approximation should be excellent and allow
to study from first principles all the low temperature anomalies 
which have been observed
in glasses. Within this approximation one just
needs to study the liquid of the centers of masses of the molecules,
which interact through the effective interaction described in
(\ref{Z_harm_resum}). Of course the interaction term coming from the $\Tr\log$
term is not easy to deal with, but still this is a
very well defined problem of liquid theory for which
precise approximation scheme should be developed.

\item Use approximations different from HNC, which may work better in the
liquid 
phase. Obviously this will depend on the interaction potential, and
a detailed study of several different types of potentials would be very 
interesting.

\item Use numerical simulation in the liquid phase in order to get 
some higher order 
coefficients of the $A$ 
expansion: these are given by higher order correlation functions which 
could be measured in simulations. 

\item Introduce resummation techniques that are more efficient than the
harmonic 
one.

\end{itemize}
Some of the previous described techniques could also  be used to understand
better the 
properties of the 
dynamical phase transition.

To summarize, our approach transforms the problem of the thermodynamics of the
glass phase into a problem of a (complicated) liquid state. We hope that
the sophisticated methods developed in liquid state theory will be brought to
bear 
on the study of glasses.

\section*{Acknowledgements} \nn
It is a pleasure to thank
David Dean and R\'emi Monasson for useful discussions. The work of MM
has been supported in part by the National Science Foundation
under Grant No. PHY94-07194.

\section*{Appendix I:  HNC closure} \nn
\label{app_HNC}
For completeness, we give here a derivation of the HNC free energy
(\ref{Ghnc}) 
for our
molecular replicated system. One could use the standard diagrammatic method 
\cite{HanMc}, 
but here we shall follow the 
'cavity' like method of Percus \cite{percus}.
We study $N$ molecules with coordinates $x_i, i \in \{ 1,..,N\} $. Each $x_i$ 
stands
 for the coordinates of all atoms in molecule $i$: $x_i=\{ x_i^a \},a \in \{ 
1,...,m\} $.
The energy of the system is given by
\be
E=\sum_{i<j} V(x_i,x_j) + \sum_i u(x_i)
\ee
where $v$ is the intermolecular potential (in our
case we would have $V(x,y)=\sum_a v(x^a-y^a)$ but we shall keep
a general $V$ in this appendix),
and the external potential $u(r)$ has been introduced for future use.

We shall need the following definitions. The one molecule density is
\be
\rho(x)=\sum_i \la \prod_a \delta (x_i^a-x^a) \ra \ ,
\ee
where the average $\la.\ra$ is with respect to the Boltzmann measure 
$\exp(-\beta E)$.
The two molecules correlation is:
\be
\rho^{(2)}(x,y)=\sum_{i \ne j} \la \prod_a \delta (x_i^a-x^a)
 \prod_b \delta (x_i^b-x^b)\ra \equiv \rho(x) g(x,y) \rho(y)
 \ee
where we have also defined the pair correlation function $g(x,y)$,
which goes to one at large (center of mass) distance. The
connected pair correlation is:
\be
h(x,y) \equiv g(x,y)-1
\ee
Elementary functional differentiation gives:
\be
{\partial \rho(x) \over \partial (-\beta u(y))}=
\rho(x) \delta(x-y) + \rho(x) h(x,y) \rho(y)
\ee
One can also introduce the direct correlation function $c(x,y)$ through:
\be
{\partial (-\beta u(x)) \over \partial \rho(y)} = {1 \over \rho(x)}
\delta(x-y) 
-c(x,y)
\ee
the direct correlation is thus related to the connected pair correlation through
the Ornstein-Zernike equation $c=(1+h\rho)^{-1}h$ which reads more explicitly:
\be
c(x,y)=h(x,y)+\int dx_1 h(x,x_1)\rho(x_1) h(x_1,y) 
+\int dx_1 dx_2h(x,x_1)\rho(x_1) h(x_1,x_2) h(x_2,y)+...
\label{ch_rel}
\ee

The idea of Percus is to compute the pair correlation by considering the one 
point density with
a molecule fixed at one point. Let us consider a problem in which we have
added 
one
extra molecule, fixed at a point $z=\{ z^1,...,z^m \} $. This extra molecule
creates an external potential $u(x)=V(x,z)$. The one point
density in the presence of this external potential, $\rho_u(x)$, is
related to the density $\rho(x)$ and pair correlation $g(x,z)$ in the absence of
an external potential through the conditional probability equation:
\be
\rho_u(x)=\rho^{(2)}(x,z)/\rho(z)= \rho(x) g(x,z)
\ee
In order to try to build a successful approximation scheme, let us
introduce two quantities $R_u(x)$ and $S_u(x)$ which we can calculate
in presence of the external potential, or when this potential is turned off 
($u=0$).
If their variations are smooth enough, one can approximate their variations
by the first order term:
\be
R_u(x) \simeq R_{u=0}(x) + \int dy {\delta R(x) \over \delta S(y)}_{u=0} \((
S_u(y)-S_{u=0}(y)\))
\label{linear}
\ee
The standard perturbation theory would be obtained by taking $R_u(x)=\rho_u(x)$
and $S_u(x)=u(x)$. However the linear truncation (\ref{linear}) can be better 
behaved
with some better choices of the functions $R$ and $S$. The HNC
 closure corresponds to taking
\cite{percus}:
\be
R_u(x) =\log \(( \rho_u(x) e^{\beta u(x)} \)) \ \ \ ; \ \ \ S_u(x)=\rho_u(x) \ .
\ee
Then we have 
\be
{\delta R(x) \over \delta S(y)} \(({u=0}\)) =c(x,y)
\ee
and the linear equation (\ref{linear}) becomes:
\be
\log g(x,z) + \beta V(x,z) = \int dy c(x,y) \rho(y) h(y,z)
\label{hnc}
\ee
Together with the inversion relation (\ref{ch_rel}), this defines a closed 
set of equations for the one and two point molecular densities which 
are the HNC closure. 
It
is easy to show that these equations express
the stationarity of the free energy functional 
$\psi[\rho,g]$ defined in (\ref{Ghnc}), 
with respect to variations of $g$.

The result for the free energy can be deduced if
we assume that:
\begin{itemize}
\item There exists a variational principle where the  free energy 
is a functional of $g$ and 
$\rho$.
\item The potential $\beta V(x)$ enters in the free energy is such a way that 
the internal energy takes the exact form $
1/2 \int dy dx \rho(x) \rho(y) g(x,y) V(x,y)$.
\item The free energy functional at $g=1$ and $v=0$, which depends only on 
$\rho$ 
is given by the 
exact form 
\be
{ \beta
 \over m} \int \prod_a d^d x^a \  \rho(x) \log { \rho(x) \over e}
\ee
\end{itemize}

These three conditions fix in a unique way the free energy functional and are 
satisfied in the 
previous approach.  Indeed the second condition 
implies that the free energy 
$\psi$ can be written as
\be
\beta \psi = \beta/2 \int dy dx \rho(x) \rho(y) g(x,y) V(x,y)+ \chi[g,\rho]
\ee
where $\chi$ does not depends explicitly on $\beta$.  
If we differentiate the previous equation with respect to $g$ we find 
\be
\beta/2 \rho(x) \rho(y) V(x,y) +{\delta \psi \over \delta g(x,y)}.
\ee
If we identify the previous equation with eq.(\ref{Ghnc}) (after
multiplication 
by $\rho(x)\rho(y)$) 
we find that the proposed free energy (eq.  (\ref{hnc})) has the same
derivative 
with respect to $g$ 
of the exact one.  Now the only ambiguity that remains in the free energy 
is its value at $g=1$ 
and $v=0$, which is fixed from the condition (3).

\section*{Appendix II: Second order small cage expansion} \nn
\label{app_ordre2}
Here we carry out the small cage expansion of the molecular HNC equations to 
second order.
We start from the HNC free energy (\ref{Ghnc}), we introduce the center of
mass 
and
relative coordinates, $x^a=X+u^a$ and $y^a=Y+u^a$,
 and we expand in the cage size $A$, using the molecular
density (\ref{rhodeu}) and the decomposition of the correlation function given
in (\ref{STdef}). 

We shall examine successively the various pieces of $2 m \beta \psi$. The
form of the  simplest piece is
deduced trivially from the constraints (\ref{constraints}):
\be
 \int dx dy \ \rho(x) \rho(y) (1-g(x,y))= \int dX dY \ (1-G(X,Y))
 \ee
(we remind that here $x$ and $y$ stand for all the molecular coordinates and are
therefore $md$-dimensional vectors, while the center of mass coordinates
$X$ and $Y$
are $d$-dimensional).

We go next to the piece involving the potential:
\be
\beta \sum_a \int dx dy \  \rho(x) \rho(y) \  v(x^a-y^a) g(x,y) \ .
\ee
We expand the potential as:
\be 
\sum_a v(x^a-y^a)= m v(X-Y) +{1 \over 2} \sum_{\mu\nu} v_{\mu\nu}(X-Y)
\sum_a (u^a_\mu-v^a_\mu)(u^a_\nu-v^a_\nu) +...
\label{expV}
\ee
and expand the correlation according to (\ref{STdef}). Thanks to the constraint
(\ref{constraints}), the term in $m v(X-Y)$ contributes exactly as:
\be
m \int dX dY \ G(X-Y) v(X-Y)
\ee
to all orders in $u,v$. The term in $ v_{\mu\nu}(X-Y)$ contributes a piece of 
order 
$A$ which is:
\be
\int dX dY  \rho(u) du  \rho(v) dv \ G(X-Y) 
 \sum_{\mu\nu} v_{\mu\nu}(X-Y)
\sum_a (u^a_\mu-v^a_\mu)(u^a_\nu-v^a_\nu)
\ee
and a piece of order $A^2$ which is:

\begin{eqnarray}
&&\int dX dY \rho(u) du  \rho(v) dv  \sum_{\mu\nu}   v_{\mu\nu}(X-Y)
\sum_a(u^a_\mu-v^a_\mu)(u^a_\nu-v^a_\nu)
\\
&&\left[ S_{\mu\nu}(X-Y) \(( \sum_b[u^b_\mu u^b_\nu+v^b_\mu v^b_\nu]-2 
K_{\mu\nu}\))
+T_{\mu\nu}(X-Y) \(( \sum_b [(u^b_\mu-v^b_\mu) (u^b_\nu- v^b_\nu)]-2 
K_{\mu\nu}\))
\right] \nonumber
\end{eqnarray}
The last piece of order $A^2$ comes from the fourth derivative of $v$ in 
(\ref{expV}):
\be
\int dX dY \rho(u) du  \rho(v) dv \ G(X-Y)
 \sum_{\mu\nu\rho\sigma} v_{\mu\nu\rho\sigma}(X-Y)\sum_a 
(u^a_\mu-v^a_\mu)(u^a_\nu-v^a_\nu)
(u^a_\rho-v^a_\rho)(u^a_\sigma-v^a_\sigma)
\ee
Notice that the use of (\ref{constraints}) allow us to find the order $A^2$ 
expression
without ever introducing the order $A^2$ term in the expansion of the pair 
correlation.
This will also be true for the other contributions below.
This strategy is crucial for keeping the computation not too big. The various 
pieces are now
easily computed using the fact that $u$ and $v$ variables are gaussian 
distributed
with the second moment given in (\ref{u2av}). We get:
\bea
\beta \sum_a\int dx dy \rho(x) \rho(y)  \  v(x^a-y^a) g(x,y)
= V  \int dX  G(X) \((m v(X) + A (m-1)  \sum_{\mu} v_{\mu\mu}(X) \))
\\
+ V A^2 (m-1) \int dX \sum_{\mu\nu} \((2 S_{\mu\nu}(X) v_{\mu\nu}(X) +
4 T_{\mu\nu}(X) v_{\mu\nu}(X) +{m-1 \over 2m} v_{\mu\mu\nu\nu}(X)
\))
\eea

We now turn to the `$g \log g$' term in the free energy $2 m \beta \psi$. 
Expanding as before,
we get:
\begin{eqnarray}
&&\int dx dy \rho(x) \rho(y) g(x,y) \log g(x,y)=
\int dX dY G(X-Y) \log G(X-Y) +
\\ \nn
&&+\int dX dY  \rho(u) du  \rho(v) dv \ {1 \over 2 
G(X-Y)}\sum_{\mu\nu\rho\sigma}
\\ \nn 
&&\[[S_{\mu\nu}(X-Y) \(( \sum_b[u^b_\mu u^b_\nu+v^b_\mu v^b_\nu]-2 
K_{\mu\nu}\))+
T_{\mu\nu}(X-Y) \(( \sum_b [(u^b_\mu-v^b_\mu) (u^b_\nu- v^b_\nu)]-2 
K_{\mu\nu}\)) \]]
\\ \nn
&&\[[S_{\rho\sigma}(X-Y) \(( \sum_b[u^b_\rho u^b_\sigma+v^b_\rho v^b_\sigma]-2 
K_{\rho\sigma}\))+
T_{\rho\sigma}(X-Y) \(( \sum_b [(u^b_\rho-v^b_\rho) (u^b_\sigma-
v^b_\sigma)]-2 
K_{\rho\sigma}\)) 
\]]
\eea
which gives after performing the gaussian $u$ and $v$ integrals:
\bea
V  \int &dX& \(( G(X) \log G(X) +{4 A^2 (m-1) \over G(X)} \sum_{\mu\nu}
\right.
\\
&&\left.
\[[ {1 \over 2} S_{\mu\nu}(X) S_{\mu\nu}(X) + S_{\mu\nu}(X) T_{\mu\nu}(X)
+T_{\mu\nu}(X) T_{\mu\nu}(X) \]] \))
\end{eqnarray}
We now study the last piece of $2 m \beta \psi$, namely the convolution term
\be
\sum_{p=3}^\infty {(-1)^p \over p} \int dx_1...dx_p \ 
\rho(x_1) h(x_1,x_2) \rho(x_2) h(x_2,x_3) ... \rho(x_p) h(x_p,x_1) \ .
\label{HNC_chain}
\ee
Here again each $x_j$ is a $md$ dimensional vector including all molecular
coordinate, which we decompose into the center of mass $X_j$ and the 
relative coordinates $u^a_j$. Therefore each piece $h(x_j,x_{j+1})$ in the
above product is expanded as:
\bea
&&h(x_j,x_{j+1})=
h(X_j-X_{j+1})
\\ \nn
&&+ \sum_{\mu\nu}\[[ S_{\mu\nu}(X_j-X_{j+1})
\(( \sum_b[u^b_{j,\mu} u^b_{j,\nu}+u^b_{j+1,\mu} u^b_{j+1,\nu}]-2 
K_{\mu\nu}\)) 
\right.
\\ 
&&\left.
+T_{\mu\nu}(X_j-X_{j+1}) \(( \sum_b [(u^b_{j,\mu}-u^b_{j+1,\mu}) 
(u^b_{j,\nu}- u^b_{j+1,\nu})]-2 K_{\mu\nu}\)) \]]
\label{hexpan}
\eea
We notice again that higher order terms do not contribute to order $A^2$. The 
second order 
terms generated by the expansion (\ref{hexpan}) when it is inserted into
(\ref{HNC_chain}) are obtained by picking up the `$h(X_j-X_{j+1})$' contribution
in all but two values of $j$. In order for the result not to vanish (because of
(\ref{constraints})), we need that this two special values of $j$ be
neighbours. 
We thus get
the following order $A^2$ contribution to the convolution term:
\bea
\sum_{p=3}^\infty &{(-1)^p } &\int dX_1...dX_p \ \rho(u_1) du_1...\rho(u_p)
du_p 
\ 
\\ \nn
&&\sum_{\mu\nu\rho\sigma}
\[[S_{\mu\nu}(X_1-X_2) \(( \sum_b[u^b_{1,\mu} u^b_{1,\nu}+
u^b_{2,\mu} u^b_{2,\nu}]-2 K_{\mu\nu}\))+
\right.
\\ \nn
&&\left.
T_{\mu\nu}(X_1-X_2) \(( \sum_b [(u^b_{1,\mu}-u^b_{2,\mu}) 
(u^b_{1,\nu}- u^b_{2,\nu})]-2 K_{\mu\nu}\)) \]]
\\ \nn
&&\[[S_{\rho\sigma}(X_1-X_2) \(( \sum_b[u^b_{1,\rho} u^b_{1,\sigma}+
u^b_{2,\rho} u^b_{2,\sigma}]-2 K_{\rho\sigma}\))+
\right.
\\ \nn
&&\left.
T_{\rho\sigma}(X_1-X_2) \(( \sum_b [(u^b_{1,\rho}-u^b_{2,\rho}) 
(u^b_{1,\sigma}- u^b_{2,\sigma})]-2 K_{\rho\sigma}\)) \]]
\\
&&h(X_3-X_4)...h(X_{p-1}-X_p)h(X_p-X_1)
\eea
After performing the gaussian $u$ and $v$ integrals, we find an 
expression in terms
of the Fourier transformed functions $h(k)$, $S_{\mu\nu}(k)$ and
$S_{\mu\nu}(k)$:
\be
2 V A^2 (m-1)
\sum_{p=3}^\infty {(-1)^p } \sum_{\mu\nu} \intk h(k)^{p-2} \((
S_{\mu\nu}(k)+ T_{\mu\nu}(k)\))^2
\ee
involving a simple geometric series.

Grouping together all the pieces of the free energy $\psi$ which we have
considered, we obtain the second order expression of the free energy
used in (\ref{def_free_ord2}-\ref{hnc2}).


\begin{thebibliography}{99}
\bi{glass_revue}
Recent reviews can be found in:  C.A. Angell, Science, 
{\bf 267}, 1924 (1995) and P.De Benedetti, `Metastable liquids', Princeton 
University
Press (1997). An introduction to the theory is: J.J\"ackle, Rep.Prog.
Phys. {\bf 49} (1986) 171.
Some introduction to the very recent
developments in connection with the spin glass
ideas is given in: G.  Parisi, {\sl Proceedings of the ACS meeting}, Orlando 
(1996),
cond-mat/9701068, {\sl Lecture given at the Sitges conference, June 1996} 
cond-mat/9701034 and {\sl 
Lectures given at the Varenna summer school} 1996, cond-mat/9705312.

\bi{gpglass}
 G. Parisi Phys.Rev.Lett. {\bf 78}(1997)4581

\bi{BK1}
 W. Kob and J.-L. Barrat, Phys.Rev.Lett. {\bf 79} (1997) 3660.

\bi{BK2}
 J.-L. Barrat and W. Kob, cond-mat/9806027.

\bi{FRAPA}
S.Franz and G.  Parisi, Phys.Rev.Lett.  {\bf 79} (1997) 2486, and 
cond-mat/9711215.

\bi{ColPar}
B.Coluzzi and G.Parisi, cond-mat/9712261.

\bi{theo}
T.M. Nieuwenhuizen, Phys.Rev.Lett. {\bf 79} (1997) 1317.

\bi{kauzman}
 A.W. Kauzman, Chem.Rev {\bf 43} (1948) 219.

\bi{AdGibbs}
G. Adams and J.H. Gibbs J.Chem.Phys {\bf 43} (1965) 139; J.H. Gibbs and E.A.
Di 
Marzio, 
J.Chem.Phys. {\bf 28} (1958) 373.

\bi{BCKM} 
J.-P.  Bouchaud, L.  Cugliandolo, J.  Kurchan, M M\'e\-zard, Physica A {\bf 
226}, 243
(1996).

\bi{KiThiWo}
 T.R. Kirkpatrick and P.G. Wolynes,  Phys. Rev. {\bf
A34}, 1045 (1986); T.R. Kirkpatrick and D. Thirumalai, Phys. Rev. Lett. {\bf 
58},
2091 (1987); T.R. Kirkpatrick and D. Thirumalai, Phys. Rev. {\bf B36}, 5388 
(1987); 
T.R. Kirkpatrick, D. Thirumalai and P.G. Wolynes,  Phys. Rev. {\bf
A40}, 1045 (1989).


\bi{GrossMez}
D.J. Gross and M. M\'ezard, Nucl. Phys. {\bf B240} (1984) 431.

\bi{corr_length} See for instance G.Parisi, cond-mat/9801034 and 
C. Donati, S.C. Glotzer, P.H. Poole, cond-mat/9811145.

\bi{nodis1}
 J.-P. Bouchaud and M. M\'ezard; J. Physique I (France) {\bf 
4} (1994) 1109.
E.  Marinari, G.  Parisi and F.  Ritort; J.  Phys.  {\bf A27} (1994) 7615; J.  
Phys.  {\bf A27} (1994) 7647.

\bi{nodis2}
P.Chandra, L.B.Ioffe and D.Sherrington, Phys. Rev. lett. {\bf 75} (1995) 713,
and cond-mat/9809417.
P.Chandra, M.V. Feigelman and L.B.Ioffe, Phys. Rev. lett. {\bf 76} (1996) 4805.

\bi{nodis3}
 E. Marinari, G. Parisi and F. Ritort, cond-mat/9410089.
 S. Franz and  J. Hertz, {\it Phys. Rev. Lett.} {\bf 74}, 2114 (1995).


\bi{remi} R. Monasson, Phys. Rev. Lett. {\bf 75} (1995) 2847.
\bi{pot} S. Franz, G. Parisi, J. Physique I {\bf 5} (1995) 1401.

\bi{MPhnc}
 M.M\'ezard and G.Parisi, J.  Phys.  A {\bf 29} 65155 (1996).

\bi{MPglass1} M.M\'ezard and G.Parisi, cond-mat/9807420.

\bi{still} F.H. Stillinger, Science {\bf 267} (1995) 1935, and references 
therein.

\bibitem{MPV} M .M\'ezard, G. Parisi and M.A. Virasoro, {\sl Spin glass theory 
and 
beyond}, World Scientific (Singapore 1987).

\bibitem{parisibook2} G. Parisi, {\sl Field Theory, Disorder and
Simulations}, World Scientific, (Singapore 1992).

\bi{kepler} An introduction to Kepler's conjecture and some description of the 
recent work by T.C.Hales which may have established the conjecture can be
found 
on the web site: 
www.math.lsa.umich.edu/~hales/countdown/.

\bibitem{crisomtap}
 A.~Crisanti and H.-J. Sommers, J. Phys. I (France) {\bf 5}, 805 (1995);
A.~Crisanti, H.~Horner and H-J~Sommers, {\it Z.Phys.} B  {\bf 92},  257 (1993)

\bi{kurparvir} 
J.~Kurchan, G.~Parisi, and M.~A. Virasoro, J. Phys. I France {\bf 3}, 1819 
(1993).

\bi{ACP} For a careful analysis of the free energy landscape see A.  Cavagna,
I. 
 Giardina and
G.  Parisi, J.  Phys.  A: Math.  Gen.  1997, {\bf 30}, 7021 and references 
therein.

\bi{MMpspin} M. M\'ezard, cond-mat/9812024, to appear in Physica A.

\bibitem{BaBuMez} A.~Barrat, R.~Burioni, and M.~M\'ezard, J. Phys. A {\bf 29}, 
L81 (1996).

\bi{cuku} L.  F.  Cugliandolo and J.Kurchan, Phys.  Rev.  Lett.  {\bf 71}, 
1 (1993).

\bi{PAK} G.  Parisi, in {\sl The Oskar Klein Centenary}, ed.  by U.  
Lindstr\"{o}m, World 
Scientific, (1995), Il nuovo cimento {\bf 16}, 939 (1994).

\bi{MCT_exp} H.Z.Cummins et al., Phys.Rev. {\bf E47} (1993) 4223.

\bi{Kob} See the review by W.Kob, cond-mat/9809268, and references therein.

\bi{POLI} L. C. E. Struik; {\it Physical aging in amorphous polymers and other
 materials} (Elsevier, Houston 1978).

\bi{BCKM_rev} 
J.-P.  Bouchaud, L.  Cugliandolo, J.  Kurchan., M M\'e\-zard, in "Spin glasses
and random fields", A.P.Young editor, Worlds Scientific 1998.

\bi{FM} S.  Franz and M.  M\'ezard Europhys.  Lett.  {\bf 26} (1994) 209;
 Physica A {\bf 210} (1994) 48.

\bi{cukusk} L.  F.  Cugliandolo and J.Kurchan, J. Phys. A {\bf 27} (1994) 5749.

\bi{gotze} 
For a review see W. Gotze, {\em Liquid, freezing and the Glass transition},
Les 
Houches 
(1989), J. P. Hansen, D. Levesque, J. Zinn-Justin editors, North Holland.

\bi{bouchaud} J.-P. Bouchaud; J. Phys. France {\bf 2}  (1992) 1705. 

\bi{BPGM} B. Coluzzi, Paolo Verrocchio , M. M\'ezard and G.Parisi, in 
preparation.

\bi{Geszti} T. Geszti, J.Phys. C {\bf 16}(1983)5805. See also 
E.Leutheusser, Phys.Rev. A {\bf 29} (1984) 2765 and U.Bendtzelius,
W.G\"otze and A. Sj\"olander, J. Phys. C{\bf 17} (1984) 5915. More
reference to the existing literature
can be found in \cite{gotze}.

\bi{goldbart} Similar order parameters
are also used in the study of randomly crosslinked macromolecules (see the 
review
by P.M. Goldbart, H.E. Castillo and A. Zippelius
Adv. Phys. 45 (1996) 393), with the important difference that the molecules are
distinguishable because of the cross links, and cannot get out of their traps.

\bi{Toulouse}
 G.Toulouse, in "Heidelberg colloquium on spin glasses", 
I. Morgenstern and L. van Hemmen eds., Springer Verlag 1983.

\bi{carparsour}
S. Caracciolo, G. Parisi, S. Patarnello and N. Sourlas, {\it
Europhys. Lett.}
{\bf 11}, 783 (1990).

\bi{CarFraPar} M.Cardenas, S. Franz and G. Parisi, cond-mat/99712099.

\bi{BouMez_extr} J.-P. Bouchaud and M.M\'ezard, J.Phys. A{\bf 30} (1997) 7337.

\bi{FRARIE} S. Franz and H. Rieger Phys.  J. Stat. Phys.  {\bf 79} 749 (1995).

\bi{MPRR} E. Marinari, G. Parisi, F. Ricci-Tersenghi, J.J. Ruiz-Lorenzo {\sl 
Violation of the 
Fluctuation Dissipation Theorem in Finite Dimensional Spin Glasses}, cond-mat 
/9710120.

\bi{INM1}
 See for instance T. Keyes, J.Phys.Chem. A {\bf 101} (1997) 2921, and
references therein.

\bi{Stratt} 
Y. Wan and R.M.Stratt, J.Chem.Phys. {\bf 100} (1994) 5123, and
references therein.

\bi{Han}
B. Bernu, Y. Hiwatari and J.P. Hansen, J.Phys. {\bf C18}, L371 (1985);
J.N. Roux, J.L. Barrat and J.P. Hansen,J.Phys. {\bf C1} 7171 (1989).

\bi{dald}
G.Daldoss, O.Pilla, G. Viliani and G. Ruocco, cond-mat/9804113.
\bi{doye}
J.P.K. Doye, M. Miller and D.J. Wales, cond-mat/9808265.

\bi{APRV} L.Angelani, G.Parisi, G.Ruocco and G.Viliani cond-mat/9803165 (1998).

\bi{nowak} E.R. Nowak, J.B. Knight, E. Ben-Naim, H.M. Jaeger and S.R. Nagel, 
Phys.
 Rev. E {\bf 57} (1998) 1971.
 

\bi{HanMc}
See for instance J.P. Hansen and I.R. Macdonald, "Theory of simple liquids", 
(Academic, London, 
1986), or H.N.V. Temperley, J.S. Rowlinson and G.S. Rushbrooke, "Physics of 
simple liquids", 
NorthHolland (Amsterdam 1968).

\bi{percus} J.K. Percus, in {\it The Equilibrium Theory of Classical Fluids},
ed. H.L. Frisch and J.L. Lebowitz (New York: Benjamin; 1964).



\end{thebibliography}
\end{document}